# Efficient Authenticated Data Structures for Graph Connectivity and Geometric Search Problems[*]

Michael T. Goodrich [†]  Roberto Tamassia [‡]  Nikos Triandopoulos [‡,§]

University of California, Irvine   Brown University   Boston University
goodrich@ics.uci.edu   rt@cs.brown.edu   nikos@cs.bu.edu

October 28, 2018

**Abstract**

Authenticated data structures provide cryptographic proofs that their answers are as accurate as the author intended, even if the data structure is being controlled by a remote untrusted host. In this paper we present efficient techniques for authenticating data structures that represent graphs and collections of geometric objects. We use a data-querying model where a data structure maintained by a trusted source is mirrored at distributed untrusted servers, called responders, with the responders answering queries made by users: when a user queries a responder, along with the answer to the issued query, he receives a cryptographic proof that allows the verification of the answer trusting only a short statement (digest) signed by the source.

We introduce the *path hash accumulator*, a new primitive based on cryptographic hashing for efficiently authenticating various properties of structured data represented as paths, including any decomposable query over sequences of elements. We show how to employ our primitive to authenticate queries about properties of paths in graphs and search queries on multi-catalogs. This allows the design of new, efficient authenticated data structures for fundamental problems on networks, such as path and connectivity queries over graphs, and complex queries on two-dimensional geometric objects, such as intersection and containment queries.

**Keywords**: authenticated data structures, data authentication, information integrity, graph connectivity, geometric searching

# Contents

# 1  Introduction

Verifying information that at first appears authentic is an often neglected task in data structure and algorithm usage. Fortunately, there is a growing literature on correctness checking that aims to rectify this omission. Following early work on program checking and certification (e.g., [8, 56, 57]), several researchers have developed efficient schemes for checking the results of various data structures (e.g., [9, 10, 11, 24, 42]), graph algorithms (e.g., [20, 35]), and geometric algorithms (e.g., [19, 43]). These schemes are directed mainly at defending the user against an inadvertent error made during implementation. In addition, these previous approaches have primarily assumed that usage is limited to a single user on an individual machine.

With the advent of Web services and Internet computing, data structures and algorithms are no longer being used just by a single user on an individual machine. Indeed, with the development of Web applications and cloud computing services, the machine responding to a user's query could be unknown to both the data-structure author and the user. More generally, it is very common that the data source and the data distributor are different entities or machines with distinct identities, and consequently, the owner of a data set does not control the data structure used to answer queries on this set. We must recognize that, although they benefit efficiency, such scenarios open the possibility that an agent hosting a data structure or an algorithm could deliberately falsify query responses to users. When the information represented by the response to a query is crucial to the target application (e.g., it has security or financial implications), such falsification could cause significant adverse consequences. We want to guard against this possibility.

In this paper we are interested in studying a new dimension in data structure and algorithm checking—how can we design sophisticated data structures and algorithms so that their responses can be verified as accurately as if they were coming from their author, even when the response is coming from an untrusted host? Examples of the kind of information we want to authenticate include dynamic documents, online catalog entries and the responses to queries in geographic information systems, financial databases, medical information systems and scientific databases. In particular, we are interested in efficiently verifying paths and connectivity information in transportation and computer networks, even when the network is changing. In addition, we are interested in verifying complex geometric queries in spatial databases, such as ray shooting, point location and range searching queries, which are used extensively in geographic information systems.

Digital signatures, used in a per-query basis, can be used to verify simple static documents, but they are inefficient for dynamic data structures. We therefore need new techniques for authenticating data structures. The main challenge in providing an integrity service in the above contexts is that the space of possible answers is much larger than the data size itself. For example, there are $O(n^2)$ different paths in a tree of $n$ nodes, and each of these paths can have $O(n)$ edges. Requiring an authenticator to digitally sign every possible response is therefore prohibitive, especially when the data is changing due to the insertion or deletion of elements in the set.

Instead, the state-of-the-art solution for this problem is *signature amortization*. Ideally, we would like our authenticator to sign just a single *digest*, i.e., a secure short description, of our entire data structure. In our work, collision-resistant hashing is the cryptographic primitive used to produce the data digest, the latter being built from the careful combination of cryptographic hashes of subsets of our data. Thus, we consider *hash-based data authentication*. The computation of the data digest and the hashes of partial data must be performed in accordance with the type of issued queries such that answer verifications can be supported *efficiently and securely*. If we can achieve such a scheme, then verifying the answer to a query in our database can be reduced to the problem of collecting the appropriate hashes of partial data that allow a user to recompute the digest of the entire structure and, further, compare it to the digest that is signed by the authenticator.

Even when we follow this approach, however, we are faced with the challenge of how to subdivide the data in a way that allows efficient assembly of partial cryptographic hashes and efficient computation of the digest of the entire structure for *any possible* query. For simple data structures, such as dictionaries, this subdivision is fairly straightforward (say, using a linear ordering and a Merkle hash tree [44]; see also [29, 46]), but for complex structures, such as graphs, geometric structures, and structures built using the fractional cascading paradigm, this subdivision method is far from obvious. For instance, there is no linear ordering among the data items of problems of the above examples upon which one could build a hash tree.

## 1.1 A Model for Authenticated Data Structures

Our authentication model involves three parties: a trusted *source*, an untrusted *responder*, and a *user*. The source holds a structured collection $S$ of objects, where we assume that a repertoire of *query operations* are defined over $S$. If $S$ is fixed over time, we say that it is *static*. Otherwise, we say that $S$ is *dynamic* and assume that a repertoire of *update operations* are defined that modify $S$.

For example, $S$ can represent a network whose nodes and edges store data items on which the following two query operations are defined: a *connectivity query* on $S$ asks whether two given nodes of $S$ are in the same connected component and a *path query* returns a path, if any, between two given nodes. We can also define update operations on $S$ that add and/or remove nodes and edges. As a second example, $S$ can be a collection of line segments in the plane forming a polygonal chain, where an *intersection query* returns all the segments intersected by a given query line. In this case we can define update operations that insert and/or remove segments.

The *responder* maintains a copy of the collection $S$ together with *structure authentication information*, which consists of hashes (of partial data) and one time-stamped—indicating its freshness—digest about $S$ signed by the source. If $S$ is dynamic, the responder receives, together with each update on $S$, some *update authentication information*, which consists of a signed time-stamped digest describing the updated state of $S$. The *user* performs queries on $S$, but instead of contacting the source directly, it queries the responder. The responder provides the user with an answer to the query together with *answer authentication information*, which yields a cryptographic proof of the validity of the answer. The answer authentication information includes the signed time-stamped digest and a collection of hashes, carefully chosen from the structure authentication information. The user verifies the answer relying solely on trusting the source's signature: from the hashes the user recomputes the digest of $S$ and accepts the answer only if the computed digest matches the verified, fresh and authentic signed digest. In terms of security, we allow the untrusted responder to be controlled by a polynomial-time adversary and require that, subject to standard cryptographic assumptions, the responder cannot successfully cheat the user, i.e., for any query no false answer can be accepted as authentic.

The data structures used by the source and the responder to store collection $S$, together with the protocols and algorithms for queries, updates, and verifications executed by the various parties, form what we call an *authenticated data structure* [29, 40, 46, 58]. In a practical deployment of an authenticated data structure, there would be various instances of geographically distributed responders. Such a distribution scheme reduces latency, allows for load balancing, and reduces the risk of denial-of-service attacks. Also, scalability is achieved by simply increasing the number of responders.

The design of authenticated data structures should address the following goals: (1) *low computational cost:* the computations performed internally by each entity (source, responder and user) should be simple and fast; also, the memory space used by the data structures supporting the computation should be as small as possible; (2) *low communication overhead:* source-to-responder communication (updates and update authentication information) and responder-to-user communication (answer and answer authentication information) should be kept as small as possible; (3) *high security:* the authenticity of answers should be verifiable with a high degree of reliability. Thus, the cost parameters measuring the performance of authenticated data structures are: 1) total storage used; 2) update cost at the source/responder; 3) query cost at the responder; 4) verification cost at the user; and 5) source-to-responder and responder-to-user communication costs.

## 1.2 Related Work

Work related to authenticated data structures was initially motivated by its applications to the issue of *certificate revocation* in public key infrastructures (see, e.g., [1, 12, 25, 36, 46]), where the underlying problem involves authenticating membership or non-membership (namely, of issued digital certificates) in dynamic sets (i.e., the set of expired, revoked or compromised certificates). Therefore, early work is mostly concerned with *authenticated dictionaries*.

The *hash tree* scheme introduced by Merkle [44] can be used to implement a static authenticated dictionary. A hash tree $T$ for a set $S$ stores cryptographic hashes of the elements of $S$ at its leaves and a value $L(v)$ at each internal node $v$, computed by hashing the concatenation of the values of the children of $v$. The authenticated dictionary for $S$ consists of the hash tree $T$ plus the signature of the value $L(r)$ stored at the root $r$ of $T$. An element $e$ is proven to belong in $S$ by reporting the values of all nodes that are siblings of the nodes lying on the path in $T$ connecting $e$ to the root. With this approach, space is linear, and the answer authentication information and the query and verification time are logarithmic in the size of the set $S$. Kocher [36] also advocates a static hash tree approach for realizing an authenticated dictionary, but simplifies somewhat the processing needed to verify non-membership in $S$, by storing intervals instead of individual elements. Other certificate revocation schemes based on variations of hash trees are described in [12, 25].

Naor and Nissim [46] use techniques that allow the dynamization of hash trees and support the insertion and deletion of elements in logarithmic time, thus implementing a dynamic authenticated dictionary. In their scheme, the source and the responder maintain identically-implemented 2–3 trees (or Seidel's randomized search trees). The update authentication information has $O(1)$ size and the answer authentication information has logarithmic size. Goodrich and Tamassia [29] present a data structure for an authenticated dictionary based on other randomized structures, skip-lists [54]. They introduce the notion of commutative hashing and show how to embed in the nodes of a skip-list a computational directed acyclic graph (DAG) of cryptographic computations based on commutative hashing. This scheme matches the asymptotic performance of the Naor-Nissim approach [46], while simplifying the details of a software architecture and an actual implementation of a dynamic authenticated dictionary as shown in [31]. An alternative skip-list based scheme is proposed in [59]. Anagnostopoulos *et al.* [2] introduce *persistent authenticated dictionaries*, where the user can issue historical queries of the type "was element $e$ in set $S$ at time $t$". Work related to persistence and historical queries appears in [39].

Additional work on authenticated dictionaries includes distributed [61] and two-party [21, 52] extensions in the recently studied *outsourced data* model, where data resides *only* at the responder, not at the source, and where typically the source itself is the *only* user issuing queries. We note that although our three-party model and the outsourced data model are related, sharing benefits and allowing for similar authentication techniques (see also [53, 60]), they individually face different challenges. For instance, more sophisticated techniques are required when no data resides at the source and fewer cryptographic techniques are applicable in a multi-user setting.

Originally introduced in [5] based on the strong RSA assumption, and later extended in [4], cryptographic accumulators constitute an alternative cryptographic primitive for authenticating membership in sets. In [13], Camenisch and Lysyanskaya introduced a dynamic extension of the RSA accumulator, which applied to our authentication model provides short membership proofs but expensive, linear, update costs. Goodrich *et al.* [30] show how to use the RSA accumulator to realize a dynamic authenticated dictionary for a set of $n$ elements with $O(1)$ answer authentication information and verification time. Their scheme allows a tradeoff between the query and update times; for example, one can balance the two times and achieve $O(\sqrt{n})$ query and update time and $O(\sqrt{n})$ update authentication information. Papamanthou *et al.* [53] further improved the update cost to $O(n^\epsilon)$ for any fixed constant $\epsilon > 0$, while keeping all other costs constant. Work on accumulators includes the support of non-membership proofs [38] and an alternative construction [48].

A first step towards the design of more general authenticated data structures (beyond dictionaries) is made by Devanbu *et al.* [18]. Using an extension of hash trees, they show how to authenticate operations *select*, *project* and *join* in a relational database. Moreover, they present an authenticated data structure for a set of multidimensional points that supports orthogonal range queries. This latter result goes beyond simple authenticated dictionaries, but it is restricted to hashing over range trees. Martel *et al.* [40] initiated a study of authenticated queries beyond tree structures and skip-lists. They consider the class of data structures such that (*i*) the links of the structure form a DAG $G$ of bounded degree and with a single source node; and (*ii*) queries on the data structure correspond to a traversal of a subdigraph of $G$ starting at the source. Such data structures can be authenticated by means of a hashing structure that digests the entire digraph $G$ into a hash value at its source, where the size of the answer authentication information and the verification time are proportional to the size of the subdigraph traversed. They showed how this general technique can be applied to authenticate static structures for pattern matching in tries and orthogonal range searching, and presented an initial treatment of authenticating fractional cascading structures, but only for range tree data structures, where catalogs are arranged as unions in a tree. Recently, a certification-based authentication framework for general queries over dynamic data was proposed in [60]. Other work includes the authentication of XML documents [7, 17] and grid searching [3].

By combining cryptographic hashing with accumulators, Nuckolls in [49] and Goodrich *et al.* in [32] present techniques for *super efficient* authentication of one-dimensional range search queries with verification costs that depend only on the answer size but not the size of the data set. Related work also appears in [47, 51] where (aggregate) signatures are employed in hash trees to authenticate queries over outsourced databases. The model used is essentially the one of authenticated data structures, but now the data sets are relational databases residing in external memory and are queried through SQL queries which are founded on one-dimensional range search. Authentication of operations on outsourced file systems is also studied in [28, 34]. Authentication models in more adversarial settings where the data source can act unreliably have been also studied: undeniable attestation by Buldas *et al.* [12], zero-knowledge sets by Micali *et al.* [45] and consistency proofs by Ostrovsky *et al.* [50] eliminate the possibility that the source produces different authenticated answers to the same query. The works in [45, 50] also consider privacy-preserving query verification.

Finally, Tamassia and Triandopoulos in [59] study the cost that is associated with authenticated data structures, deriving the first lower bounds for set-membership using cryptographic hashing. They show that any hash-based authenticated dictionary of size $n$ has authentication overheads that are at least logarithmic in $n$ in the worst case, even when the structure authentication information has $O(n^{1-\epsilon})$ size, $\epsilon > 0$, i.e., even when that many data digests are totally signed by the source. Additional comprehensive studies on the cost of authenticated data structures have been presented by Li *et al.* [37] and Goodrich *et al.* [27] on B-trees and skip-lists respectively.

## 1.3 Our Contributions

In this paper, we present general techniques for building authenticated data structures for a broad class of query problems on graphs and geometric objects. First, we describe an authenticated data structure that represents a general graph $G$ and supports the authentication of a large set of graph queries. Through the authentication of a generic abstract type of queries related to properties about paths in $G$, we provide authentication for a large number of query problems on general graphs, including the following queries, where $v$, $w$ are nodes of graph $G$: areConnected$(v, w)$, areBiconnected$(v, w)$, areTriconnected$(v, w)$, reporting respectively whether $v$ and $w$ are in same connected, biconnected and triconnected component, as well as path$(v, w)$ and pathLength$(v, w)$, reporting respectively the nodes and the length of a path connecting $v$ to $w$. We also support efficient update operations that involve insertions of nodes and edges in $G$. Our data structure uses linear space and supports connectivity queries and update operations in $O(\log n)$ time and path queries in $O(\log n + k)$ time, where $k$ is the length of the path reported. The update authentication information has $O(1)$ size. The size of the answer authentication information and the verification time are each $O(\log n)$ for connectivity, biconnectivity and triconnectivity queries and $O(\log n + k)$ for path queries. If the graph is planar, the data structure is fully dynamic and supports arbitrary series of intermixed insertions and deletions of nodes/edges. For general graphs, the data structure supports insertions but not deletions. These results have applications to the authentication of network and file management systems.

In addition, we address several geometric search problems, showing how to authenticate the full, general version of the powerful *fractional cascading* technique [14]. Our authentication technique provides a general framework capable to authenticate any data structure built using the fractional cascading technique. In particular, we can efficiently authenticate any query for the *iterative search problem*, where we have a collection of $k$ dictionaries of total size $n$ stored at the nodes of a connected graph and we want to search for an element in each dictionary in a subgraph of this graph. Fractional cascading yields a data structure with linear-space that supports iterative search queries in $O(\log n + k)$ time. This is better than the $O(k \log n)$ time of separate searches and the $O(kn)$ space that results by searching in a merged master dictionary. A number of fundamental two-dimensional geometric searching problems can be solved with data structures based on fractional cascading [15]. These problems include: *line intersection* and *ray shooting queries* on a polygon $P$, to report the edges and respectively the first edge of $P$ intersected by a query line, *point location* on a planar subdivision, to report the region containing a query point, *orthogonal range search* on a set of points in $\mathbf{R}^2$, to report the points inside a query rectangle, as well as *orthogonal point enclosure* and *orthogonal intersection queries* on a set of rectangles, to report those that contain a query point and respectively are intersected by a query rectangle. We show that our authenticated fractional cascading data structure can be extended to yield efficient authenticated data structures for all the above problems. Our authentication schemes have applications to database management and geographic information systems.

Our authentication schemes are based on a new authentication primitive, the *path hash accumulator*, an efficient hash-based structure for authenticating various properties of structured data represented as paths. In particular, path hash accumulator supports the authentication of any decomposable query over data items stored as paths or over sequences of elements. By building on our new primitive we achieve efficiency and modularity: our schemes can be easily analyzed in terms of complexity and security and are simple to implement. Our new primitive can be used as a general-purpose authentication tool in the design of more complex authenticated data structures.

Based on widely-used cryptographic primitives, such as collision-resistant hashing and digital signatures, are scheme are practical and secure under standard hardness assumptions.

**Paper Structure.** Our paper is organized as follows. In Section 2 we define our data authentication model, state our cryptographic assumptions and present the general authentication technique used in our work. In Section 3 we present the path hash accumulator, an authentication scheme for verifying properties on sequences of elements, or more generally properties on paths. In Section 4 we present our authenticated data structures for various path and connectivity queries on graphs. In Section 5 we present the authentication of the fractional cascading algorithmic paradigm, which leads to the authentication of various geometric data structures. We conclude in Section 6.

## 2 Hash-Based Data Authentication

In this section, we present the general cryptographic technique that is used in our authenticated data structures and discuss security issues related to our authentication model.

Let $S = \{e_1, e_2, \ldots, e_n\}$ be a data set owned by the source and let $\mathcal{Q}$ be the set of all possible queries that a user can issue about $S$. In an extreme solution, the source can just digitally sign the answer to every possible query $q \in \mathcal{Q}$. Since $\mathcal{Q}$ can be infinitely large, this solution is almost always not viable. Thus, directly applying well-known message authentication techniques for data authentication (e.g., signing whatever piece of information needs authentication) is not suitable *per se*, and additional machinery is needed. Alternatively, the source can sign a set $\mathcal{C}(S) = \{s_1, s_2, \ldots, s_c\}$ of "statements" or, in fact, relations over elements in $S$ that completely describe data set $S$, meaning that each query $q \in \mathcal{Q}$ can be answered and validated by providing some minimal subset $A(q) \subseteq \mathcal{C}(S)$ of such statements to the user and proving $A(q)$ to be authentic, where $c$ is a quantity that typically depends on $n$. In particular, the source can sign all statements of $\mathcal{C}(S)$ and provide them to the responder, so that they are appropriately forwarded to the user along with the answer to an issued query. Set $\mathcal{C}(S)$ must be chosen the minimal possible one. For instance, if $S$ represents an ordered sequence of $n$ elements and $\mathcal{Q}$ is the set of all one dimensional range queries about $S$ (i.e., "report elements of $S$ in range $[x_1, x_2]$"), then $\mathcal{C}(S) = \{s_1, s_2, \ldots, s_c\}$ could consist of statements of the form "$e_{i+1}$ is the successor of $e_i$ in $S$", where $c = n + 1$, since an ordered sequence of $n$ elements defines exactly $n + 1$ relations of the form $(e_i, e_{i+1})$, where $e_{i+1}$ is the successor of $e_i$.

Our approach is *signature amortization*, which constitutes the state-of-the-art solution for data authentication and is common in concept in all works on authenticated data structures [40, 46, 58]. The idea is to compute (at the source) and maintain (at the responder) a *digest* of the data set $S$, that is, a short cryptographic description of $S$, and to have this digest be the *only* statement signed by the source. Upon a query $q$, along with the answer, the user is provided by the responder with this signed digest and with some auxiliary information (i.e., a proof) that is sufficient for verifying the answer. This verification step at the user amounts to using the answer and the auxiliary information for locally computing the digest of $S$, and comparing this against the signed and verified digest (the one directly provided by the responder). Once the two digests have been checked by the user to be identical, the user has a cryptographic proof about the correctness of the answer. This is the case because the signed digest has been constructed (by the source) in such a way so that it carries certain cryptographic properties and encodes certain structural properties of $S$ that allow the secure transmission of trust from the signed (and verified) digest to the entire data set $S$, as well as, for any query $q$, the efficient verification of the corresponding answer. In this way, one signature over the digest of the data set is amortized over all queries in $\mathcal{Q}$ on $S$.

In this work, we consider the case where the data digest is computed using *cryptographic collision-resistant hashing* (e.g., the SHA family of functions) and call this technique *hash-based data authentication*. Note that the use of accumulators (e.g., as in [13, 30, 53]) to produce the digest constitutes an alternative, though computationally significantly more expensive, approach. We next describe more formally our approach, starting by the used cryptographic primitives.

## 2.1 Cryptographic Primitives

Before presenting the cryptographic primitives that are used in our work, we introduce some notation. By $y \leftarrow A(x)$, we denote that $y$ was obtained by running algorithm $A$ on input $x$. If $A$ is probabilistic, then $y$ is a random variable. If $S$ is a finite set, then $y \leftarrow S$ denotes that $y$ was chosen from $S$ uniformly at random. By $A^O(\cdot)$, we denote an algorithm $A$ that has *oracle access* to $O$, i.e., makes queries to an oracle $O$. By $Q = Q(A^O(x)) \leftarrow A^O(x)$ we denote the contents of the query tape once $A$ terminates, with oracle $O$ and input $x$. By $(q, a) \in Q$ we denote the event that $A$ issued query $q$ and $O$ answered $a$. If $b$ is a boolean function, by $(y \leftarrow A(x) : b(y))$, we denote the event that $b(y)$ is true after $y$ was generated by running $A$ on input $x$. Finally, a function $\nu : \mathbb{N} \to \mathbb{R}$ is *negligible* if for every positive polynomial $p(\cdot)$ and for sufficiently large $k$, $\nu(k) < \frac{1}{p(k)}$.

To realize signature amortization in our authentication model, the basis of trust is the postulate that the user trusts the data source. In the public-key cryptographic model this is expressed by means of a *digital signature scheme*. The user knows the public key of the source and trusts that anything signed under the corresponding private key is authentic. For completeness we present the standard definition of the signature-scheme primitive as in [26]. Schemes that satisfy the following security requirement are said to be secure against *adaptive chosen-message attack*.

**Definition 2.1** (Signature Scheme). *Probabilistic polynomial-time algorithms* $(G(\cdot), \mathsf{Sign}_{(\cdot)}(\cdot), \mathsf{Verify}_{(\cdot)}(\cdot, \cdot))$, *where $G$ is the key generation algorithm,* $\mathsf{Sign}$ *is the signing algorithm, and* $\mathsf{Verify}$ *the verification algorithm, constitute a digital signature scheme for a family (indexed by the public key $PK$) of message spaces* $\mathcal{M}_{(\cdot)}$ *if the following two hold:*

**Correctness** *If a message $m$ is in the message space for a given public key $PK$, and $SK$ is the corresponding secret key, then the output of $\mathsf{Sign}_{SK}(m)$ will always be accepted by the verification algorithm $\mathsf{Verify}_{PK}$. More formally, for all values of $m$ and security parameter $k$:*

$$\Pr[(PK, SK) \leftarrow G(1^k); \sigma \leftarrow \mathsf{Sign}_{SK}(m) \ : \ m \leftarrow \mathcal{M}_{PK} \wedge \neg\mathsf{Verify}_{PK}(m, \sigma)] = 0.$$

**Security** *Even if an adversary has oracle access to the signing algorithm that provides signatures on messages of the adversary's choice, the adversary cannot create a valid signature on a* *message not explicitly queried.* More formally, for all families of probabilistic polynomial-time oracle Turing machines $\{A_k^{(\cdot)}\}$, there exists a negligible function $\nu(k)$ such that

$$\Pr[(PK, SK) \leftarrow G(1^k); (Q(A_k^{\mathsf{Sign}_{SK}(\cdot)}(1^k)), m, \sigma) \leftarrow A_k^{\mathsf{Sign}_{SK}(\cdot)}(1^k) :$$
$$\mathsf{Verify}_{PK}(m, \sigma) = 1 \wedge \neg(\exists \sigma' \mid (m, \sigma') \in Q)] = \nu(k).$$

Hash-based data authentication employs a *cryptographic hash function* $h$ to produce the digest of the data set. Function $h$ operates on a variable-length message $M$ producing a hash value $h(M)$ of short and fixed length. Moreover, function $h$ is called *collision-resistant* if it is computationally hard to find two different strings $x \neq y$ that hash to the same value, i.e., form a collision $h(x) = h(y)$. For completeness, we give a standard definition of a family of collision-resistant hash functions.

**Definition 2.2** (Collision-resistant Hash Function). *Let $\mathcal{H}$ be a probabilistic polynomial-time algorithm that, on input $1^k$, outputs an algorithm $h : \{0,1\}^* \mapsto \{0,1\}^k$. Then $\mathcal{H}$ defines a family of collision-resistant hash functions if:*

**Efficiency** *For all $h \in \mathcal{H}(1^k)$, for all $x \in \{0,1\}^*$, it takes polynomial time in $k + |x|$ to compute $h(x)$.*

**Collision-resistance** *For all families of probabilistic polynomial-time Turing machines $\{A_k\}$, there exists a negligible function $\nu(k)$ such that*

$$\Pr[h \leftarrow \mathcal{H}(1^k); (x_1, x_2) \leftarrow A_k(h) \ : \ x_1 \neq x_2 \wedge h(x_1) = h(x_2)] = \nu(k).$$

## 2.2 Hashing Scheme and Authentication Framework

We now describe the general framework that our authentication techniques are based on. Working in the public-key cryptographic model and given a security parameter, a signature scheme and a collision-resistant hash function $h$ are available for use to all parties (source, responder, user). Thus, there is always available information that allows the user to validate a signature produced by the source.

To achieve signature amortization, what is signed by the source is a digest $d$ of the data set $S = \{e_1, \ldots, e_n\}$ the source owns. In our authentication schemes, the collision-resistant hash function $h$ is used to produce this digest. To achieve this, we assume some well-defined binary representation for any data element $e$ of $S$, so that $h$ can operate on $e$ to produce hash value $h(e)$. Also, we assume that rules have been defined so that $h$ can operate over any finite sequence of elements, i.e., $h(e_{i_1}, \ldots, e_{i_k})$ represents a hash value computed over $k$ elements $e_{i_1}, \ldots, e_{i_k}$ of $S$. For instance, $h(e_{i_1}, \ldots, e_{i_k})$ can denote that $h$ operates on the concatenation $e_{i_1} \| \ldots \| e_{i_k}$ of $e_{i_1}, \ldots, e_{i_k}$, or on the concatenation $h(e_{i_1}) \| \ldots \| h(e_{i_k})$ of their hashes; in both cases, $h$ operates on a binary string $\sigma$, where the cost of the operation of $h$ on $\sigma$ is proportional to the length $|\sigma|$ (in particular, this is true for the SHA family of hash functions; see also [59]).

We next define *hashing schemes*, our general approach for computing a digest $d$ in a systematic way over set $S$. In particular, $d$ is computed by means of a single-sink directed acyclic graph defined over $S$, whose nodes are associated with elements in $S$ and are labeled with hash values.

**Definition 2.3** (Hashing Scheme). *Let $S = \{e_1, \ldots, e_n\}$ be a data set and $G$ be a single-sink directed acyclic graph. A hashing scheme for $S$ using $G$ is a node-labeling in $G$ created as follows. Sequences of data elements of $S$ are associated with nodes of $G$, where each sequence has constant size with respect to $n$, and each node $u \in G$ is assigned a hash value or* label $L(u)$, *such that:*

- *if $u$ is a source node of $G$, and $(e_{u_1}, \ldots, e_{u_m})$ is the sequence of elements of $S$ associated with node $u$, where $m \geq 0$ is some constant integer, then*

$$L(u) = h(e_{u_1}, \ldots, e_{u_m}); \text{ otherwise}$$

- *if $u$ is a non-source node of $G$, $(z_1, u), \ldots, (z_k, u)$ are the incoming edges of $u$ in $G$, and $(e_{u_1}, \ldots, e_{u_m})$ is the sequence of elements of $S$ associated with node $u$, where $k, m \geq 0$ are some constant integers, then*

$$L(u) = h(e_{u_1}, \ldots, e_{u_m}, L(z_1), \ldots, L(z_k)).$$

*If $t$ is the sink node of $G$, the* digest *of $S$ using this hashing scheme is label $L(t)$.*

Often, by the term hashing scheme we refer not only to the node-labeling of $G$ but rather to the augmented graph $G$, including the association between data elements and graph nodes and the hash values. Observe that we explicitly require that each node of $G$ is associated with a constant number of data elements, and that we also implicitly assume that nodes in $G$ have constant in-degree, i.e., the number of incoming edges of any node is constant. Note that as digest $d$ of $S$ is simply a hash value, it has short length. In general, $G$ is defined in accordance with the data structure used to answer queries on $S$, but without necessarily exactly coinciding with it.

Given a data set $S$ and a signature scheme, signature amortization is implemented using the following *authentication scheme.* Using a hashing scheme, the data source produces a digest $d$ of $S$ according to a specific query type $\mathcal{Q}$, and using a signature scheme, the source digitally signs $d$. This signed digest is the core trusted component for authenticating queries: the answer to a query $q \in \mathcal{Q}$ is authenticated by being checked against the validity of the signed $d$. This is achieved by the collision-resistant property of $h$ and the way digest $d$ has been computed through the hashing scheme $G$. In particular, $G$ is constructed such that it expresses structural information about $S$ that corresponds to a set of relations $\mathcal{C}(S)$ capable of verifying the answers to queries in $\mathcal{Q}$. When the signed digest $d$ is verified to be authentic, it is the collision-resistance property of $h$ that transmits trust to the data elements of $S$—from the authentic $d$ and through the graph $G$—essentially, authenticating the set $\mathcal{C}(S)$. This authentic information is finally used to check the validity of the provided answer. The hashing scheme should be designed so that, depending on the query type $\mathcal{Q}$, checking the validity of answers can be performed *efficiently, correctly and securely,* independently of the specific query $q$.

For dynamic data evolving over time, new digests are computed and signed by the source. To avoid *replay attacks* launched by the responder, that is, attempts for answer verification subject to old, invalid data digests (that can be easily cached by the responder), the technique of *time-stamping* is used, as it was introduced in [46]. A digest is signed after a time-stamp is appended to it, which is used by the user to check the freshness of the signature on the digest. A verifiable answer is finally accepted only if it corresponds to a fresh signature, that is, only if the time-stamp is recent (according to some convention depending on the higher level application). Accordingly, the source periodically resigns the current digest, even if no changes occur in the data set.

Overall, our authentication techniques are based on the following general protocol. The source and the responder store identical copies[1] of the data structure representing $S$ and maintain the same hashing scheme $G$ on $S$. The source periodically signs the digest $d$ of $S$ together with a fresh time-stamp and sends this signed time-stamped digest to the responder. When updates occur on $S$, these are sent to the responder together with the new signed time-stamped digest. Note that in this setting (signature amortization and hash-based authentication), the update authentication information has $O(1)$ size and the structure authentication information consists of $G$ (hash values).

---

[1] When randomized data structures are in consideration, identical copies can be still maintained by having the source and the responder sharing the same randomness seed.

When the user poses a query, the responder returns to the user the answer along with the answer authentication information, i.e., it returns (*i*) the answer to the query, (*ii*) the signed time-stamped digest of $S$ and (*iii*) a proof, consisting of a small collection of labels (hash values) or data elements from the hashing scheme $G$ that allows the recomputation of the digest and the semantic verification of the answer. The user validates the answer by recomputing the digest (and checking its correctness as it is expressed by the hashing scheme and the corresponding set $\mathcal{C}(S)$), checking that it is equal to the signed one and verifying the signature on the digest. Accordingly, the user either verifies the authenticity of the answer and *accepts* it as authentic, or otherwise, the user *rejects* the answer. The total time spent for this process is called the *answer verification time*.

Note that, at the user side, the verification algorithm operates on the three inputs: the answer, the proof and signed digest. The answer and the proof are used to recompute the digest. In doing this, the user employs the collision-resistant hash function $h$ in combination with the hashing scheme $G$. Both $h$ and the structure of $G$ are assumed to be available to the user as part of the public key. Alternatively, we can think the subgraph of $G$ used by the user to be part of the proof. The hashing scheme $G$ encodes set $\mathcal{C}(S)$ and is the means by which the user associates the answer with the proof and the issued query $q$ and finally verifies the answer. Note that our authentication framework is appropriate for any type of queries $\mathcal{Q}$ and depends on the hashing scheme in use (indeed, Definition 2.3 imposes no restriction on the exact structure of the used DAGs). As we discuss next, the design of efficient and secure authenticated data structures is based on the correct and careful definition of a hashing scheme $G$ in accordance with the type of queries $\mathcal{Q}$.

## 2.3 Security

Finally, we discuss the security requirement for any authentication scheme and how it is achieved using the above authentication method. Here, we present a security definition appropriate for the model of authenticated data structures. A more formal security definition is available in [60].

Starting from the basis that the user trusts the data source but not the responder, it is the responder that can act adversarially. We first assume that the responder always participates in the three-party protocol, i.e., it communicates with the source and the user, as the protocol dictates. Thus, we do not consider denial-of-service attacks; they do not form an authentication attack but rather a data communication threat. Actually, although practically nothing prevents the responder from denying to participate in the protocols (e.g., refusing to respond to a user's query), in principle nothing prevents the user from redirecting the query to another responder. Indeed, a practical deployment of authenticated data structures utilizes responders as geographically distributed— widely spread in a network—mirror sites of the source, thus, the user can contact more than one responders. Additionally, if a responder is a service provider, denial-of-service attacks can be prevented using some form of penalties applied to non-cooperative responders.

However, a responder can try to cheat, by not providing the correct answer to a query but attempting to forge a fake proof for a false answer. We model this scenario by assuming that the responder is controlled by a computationally bounded (polynomial-time) adversary $\mathcal{A}$. The adversary performs a type of attack that is similar to an adaptive chosen-message attack for signature schemes. That is, $\mathcal{A}$ has oracle access to the authentication technique and possesses the signed digest (using a particular hashing scheme) of any data set $S'$ of his choice. Then, given a particular query $q \in \mathcal{Q}$ for a data set $S$, the goal of the adversary is to construct a false answer and a fake proof for this query that passes the verification check performed at the user. Accordingly, the security requirement that the authentication scheme of any authenticated data structure should satisfy is as follows: given any query by a user, no polynomial-time responder can reply with a pair of answer and an associated proof, such that both the answer is not correct and the user (incorrectly) verifies the authenticity of the answer and accepts it. More formally, we require the following property:

**Definition 2.4** (Security Requirement). *An authenticated scheme for an authenticated data structure is* secure, *if for any query issued by a user, no polynomial-time adversary $\mathcal{A}$—controlling the responder that answers the query and having oracle access to the authentication scheme—has non-negligible in the security parameter advantage in causing a user to accept, i.e., to verify as correct, an incorrect answer.*

For hash-based data authentication, our techniques follow the standard "hash and sign" authentication paradigm, and the above property is achieved by relying on the security properties of signatures and collision-resistant hashing. That is, if the hashing scheme is carefully designed such that it encodes statements about the data set $S$ (set $\mathcal{C}(\mathcal{S})$) that allow answer verification, then the authentication scheme is secure: using standard arguments, any attack against the scheme is reduced to an attack either on the signature scheme or on the collision-resistant hash function. Thus, the security of an authentication scheme is fully characterized by the hashing scheme.

We close our discussion by noting that for search query problems, where the answer to a query is a subset of the data elements in $S = \{e_1, \ldots, e_n\}$, we can further characterize the properties that the underlying hashing scheme should satisfy. Let $a(q) = \{e_{a_1}, \ldots, e_{a_l}\} \subseteq S$ be the unique answer of a search query $q \in \mathcal{Q}$ for a data set $S$, where $0 \leq l \leq n$ is the size of the answer. If $a'(q) = \{e_{a'_1}, \ldots, e_{a'_k}\}$ is the answer of size $k$, $0 \leq k \leq n$, given by the responder to the user, then the hashing scheme $G$ should be chosen such that it can be used to check whether $a'(q)$ is correct. And in this case, answer $a'(q)$ is said to be *correct*, if it contains: (1) only elements that satisfy the query parameters of $q$ and (2) all elements that satisfy the query parameters of $q$, i.e., if $a'(q) = a(q)$ (see also [40]).

## 3 Authenticated Path Properties

We now present the *path hash accumulator*, our first authentication scheme that provides authentication of a general class of queries on a sequence $S = (e_1, e_2, \ldots, e_n)$. That is, here the data set is an ordered collection of $n$ elements, where the notion of *predecessor* and *successor* are defined on elements of $S$ and the notion of *first* and *last* are defined on $S$. Our path hash accumulator will serve as a primitive authentication tool used in the rest of the paper (Sections 4 and 5). We start by introducing some notation related to paths, the central technical concept in our work.

### 3.1 Paths and Path Properties

An abstract notion of a *path* is used to represent sequence $S$. We use and extend the notation in [16]. A *path* consists of one or more *nodes* and is directed, in accordance with the need to capture the predecessor-successor relationship. Also, paths can be joined to form *concatenation paths* and they can define *subpaths*. More formally:

**Definition 3.1.** *A* path *$p$ is an ordered sequence of one or more nodes. The first and last nodes of a path $p$ are called the* head *and* tail *of $p$, and are denoted as $head(p)$ and $tail(p)$. A path is considered to have a direction: each node is connected to its successor by a directed edge. If $p'$ and $p''$ are paths, the* concatenation *$p = p' \| p''$ is a path formed by adding a directed edge from $tail(p')$ to $head(p'')$. A* subpath *$\bar{p}(v, u) = \bar{p}$ of a path $p$ is the path consisting of the collection of consecutive nodes $v, w_1, \ldots, w_l, u$ of $p$ with $head(\bar{p}) = v$, $tail(\bar{p}) = u$, where $w_1, \ldots, w_l$, $l \geq 0$, are the intermediate nodes of $\bar{p}$ that lie between $v$ and $u$, and where $v = u$ is possible whenever $l = 0$.*

The data set $S$ is associated with a path through the notion of *node attributes* and *node properties*, values that are stored in the nodes of the path. Similarly, *path attributes* and *path properties* extend node attributes and node properties to a collection of consecutive nodes, i.e., to paths.

**Definition 3.2.** *A* node attribute *$N(v)$ of node $v$ is a value related to and stored at $v$. $N(v)$ can assume arbitrary values and occupies only $O(1)$ storage. A* node property *$\mathcal{N}(v)$ of node $v$ is a sequence $N_1(v), \ldots, N_r(v)$ of node attributes, where $r$ is a constant. For a node $v$, we require that $v$ is included in any node property $\mathcal{N}(v)$ of $v$ as a node attribute of $v$. A* path attribute *$P(p)$ of path $p$ is a value that is related to $p$ and occupies only $O(1)$ storage. A* path property *$\mathcal{P}(p)$ of $p$ is a sequence of path attributes $P_1(p), \ldots, P_s(p)$, where $s$ is a constant. For a path $p$, we require that $head(p)$ and $tail(p)$ are included in any path property $\mathcal{P}(p)$ of $p$ as path attributes of $p$.*

A path property is a sequence of path attributes, that is, a sequence of values that relate to the path, in particular, values that depend on the data stored at the nodes of the path as node properties and possibly on the structural properties of the path (e.g., path size, node ordering etc.). Therefore, path properties depend on the corresponding node properties defined over the path. Moreover, the definition of path properties (and path attributes) is naturally extended to the case where subpaths of paths are considered. Accordingly, we can view a path property as a *mapping* $\mathcal{P}$ from paths (and, actually, node properties of their nodes and the path structure) to sequences of values. That is, path property $\mathcal{P}(p)$ of path $p$ is a mapping from $p$ to a sequence of values related to $p$. Thus, $\mathcal{P}(\cdot)$ can be treated as a function. Also, note that a node property $\mathcal{N}(u)$ can be viewed as a corresponding path property $\mathcal{P}(\bar{p}(u, u))$ and vice versa.

We are interested in path properties that satisfy the *concatenation criterion*.

**Definition 3.3.** *Let $p$ be a path and let $p'$ and $p''$ be any subpaths of $p$ such that $p = p' \| p''$. A path property $\mathcal{P}$ satisfies the* concatenation criterion *if $\mathcal{P}(p) = \mathcal{F}(\mathcal{P}(p'), \mathcal{P}(p''))$, where $\mathcal{F}$ is a function defined on pairs of sequences of values (path attributes) that can be computed in $O(1)$ time. Function $\mathcal{F}$ is called the* concatenation function *of $\mathcal{P}$.*

In other words, a path property satisfies the concatenation criterion when this path property evaluated for a path $p$ can be computed in constant time given the corresponding path properties of any two subpaths of $p$. Thus, a path property satisfies the concatenation criterion when its corresponding concatenation function admits a computational evaluation that is inherently *associative*.

We wish to be able to locate nodes of a path that are of our interest. This is achieved by a *node selection query* by means of a *path selection function*. A *path selection query* extends a node selection query for locating subpaths using a *path advance function*.

**Definition 3.4.** *Let $\mathcal{P}$ be a path property that satisfies the concatenation criterion. Given a path $p$ and a query argument $q$, a* node selection query *$Q_\mathcal{N}$ maps $p$ into a node $v = Q_\mathcal{N}(p, q)$ of $p$. A node selection query is always associated with some* path selection function. *Given that $p = p' \| p''$,* a path selection function $\sigma(p, q)$ for $Q_\mathcal{N}$ determines in $O(1)$ time whether $v$ is in $p'$ or $p''$ using $q$ and values $\mathcal{P}(p')$ and $\mathcal{P}(p'')$.

**Definition 3.5.** *Let $\mathcal{P}$ be a path property that satisfies the concatenation criterion. Given a path $p$ and a query argument $q$, a path selection query $Q_\mathcal{P}$ maps $p$ into a subpath $\bar{p} = Q_\mathcal{P}(p, q)$ of $p$. A path selection query is always associated with some path advance function. Given that $p = p' \| p''$, a path advance function $\alpha(p, q)$ for $Q_\mathcal{P}$, using values $\mathcal{P}(p')$ and $\mathcal{P}(p'')$, returns in $O(1)$ time the subpath(s) among $p'$, $p''$ (possibly none) for which the query argument $q$ holds.*

Let $p$ be a path, $\mathcal{P}$ be any path property that satisfies the concatenation criterion and let $\mathcal{N}$ be any node property. We are interested in authenticating the following query operations on $p$:

- property(subpath $\bar{p}(v, u)$): report the value of path property $\mathcal{P}$ for subpath $\bar{p}(v, u)$ of $p$ ($\bar{p}$ may be equal to $p$);

- property(node $v$): report the value of node property $\mathcal{N}$ for node $v$;

- locate(path $p$, path selection function $\sigma$, argument $q$): find node $v = Q_\mathcal{N}(p, q)$ of $p$ returned by the node selection query $Q_\mathcal{N}$ expressed by the path selection function $\sigma$;

- subpath(path $p$, path advance function $\alpha$, argument $q$): find the subpath $\bar{p} = Q_\mathcal{P}(p, q)$ of $p$ returned by the path selection query $Q_\mathcal{P}$ expressed by the path advance function $\alpha$.

That is, we are interested in authenticating the "path property" query property($\cdot$) that returns the corresponding node or path property of its argument, and "search" queries locate($\cdot$), subpath($\cdot$) that, having a reverse role and using the path or node properties, return a path or node.

## 3.2 Path Hash Accumulator

We now present our first authentication scheme for the above query operations on paths, discussing the details of the path representation and its associated hashing scheme. Let $\mathcal{P}$ and $\mathcal{N}$ be the path property satisfying the concatenation criterion and the node property of our interest.

We represent a path $p$ as a balanced binary tree $T(p)$, where each leaf of $T(p)$ represents a node of $p$, so that the left-to-right ordering of the leaves corresponds to $p$. (We do not distinguish between path nodes and corresponding tree leaves.) An internal node $v$ of $T(p)$ represents the subpath $p(v)$ of $p$ that corresponds to the leaves of the subtree rooted at $v$. Each leaf $u$ stores the corresponding node property $\mathcal{N}(u)$ and each internal node $v$ stores the corresponding path property $\mathcal{P}(p(v))$. For simplicity, we denote the path property $\mathcal{P}(p(u))$ of the subpath defined by node $u$ as $\mathcal{P}(u)$.

The *path hash accumulator* for path $p$ is the hashing scheme for the node and path properties of $p$ using a DAG that is induced by tree $T(p)$. Specifically, consider the data set consisting of: (1) for each leaf node $v$ of $T(p)$, the node property $\mathcal{N}(v)$ and (2) for each internal node $u$ of $T(p)$, the path property $\mathcal{P}(u)$ of the subpath $p(u)$ associated with the leaves in the subtree rooted at $u$. Let $G$ be the DAG obtained from $T(p)$ by directing each edge towards the parent node, and let $h$ be a collision-resistant hash function. Also, for a node $v$ of $p$, let $pred(v)$ and $succ(v)$ denote the predecessor and the successor of $v$ in $p$, respectively. In particular, $pred(head(p))$ and $succ(tail(p))$ are some special $\bot$ (nil) values. Then given $h$, the hashing scheme of $p$ (actually, of the above data set) using $G$ is defined by computing a label $L(u)$ for each node $u$ of $T(p)$ as follows:

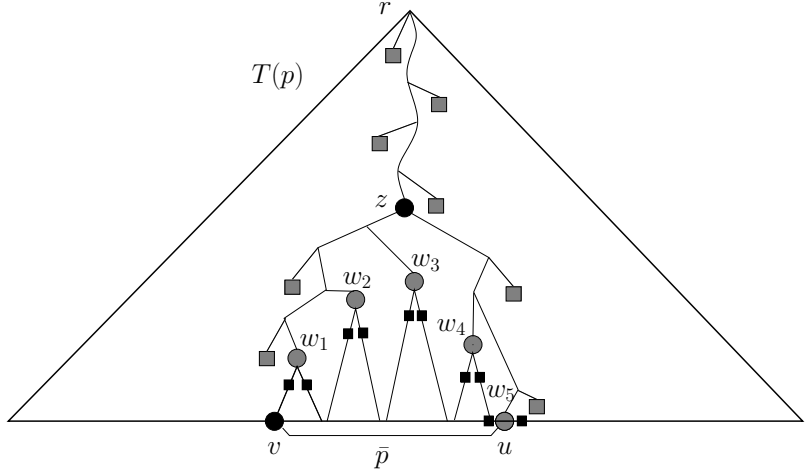

Figure 1: Answer authentication information for property($\bar{p}(v,u)$) consisting of: (1) the properties of allocation nodes $w_1, \ldots, w_5$ (grey circle nodes); (2) the labels of the children of the allocation nodes, if these children exist, or else the node properties of their neighboring nodes in $p$ (black square nodes); (3) the labels and properties of sibling nodes of nodes in the leaf-to-root paths from $v$ and $u$ up to $r$ that are not allocation nodes (grey square nodes).

- If $u$ is a source vertex of $G$, i.e., a leaf of $T(p)$, then

$$L(u) = h(\mathcal{N}(pred(u)), \mathcal{N}(u), \mathcal{N}(succ(u))), \qquad (1)$$

  where, by convention, $\mathcal{N}(\bot)$ takes on a special, fixed, hash value;

- If $u$ is a non source vertex of $G$ and $(z_1, u)$ and $(z_2, u)$ are edges of $G$, then

$$L(u) = h(\mathcal{P}(u), L(z_1), L(z_2)). \qquad (2)$$

The digest of the above data set is the label $L(r)$ of the sink $r$ of $G$ (i.e., $r$ is the root of $T(p)$) and is called the *path hash accumulation* of path $p$.

We are interested in supporting the following two update operations on paths. Operation concatenate(path $p'$, path $p''$) joins paths $p'$ and $p''$ to the concatenation path $p = p' \| p''$ and operation split(node $v$) splits the path $p$ that contains $v$ in two subpaths $p'$ and $p''$ such that $p = p' \| p''$ and $v = head(p'')$. These primitive operations are useful to support more complex update operations of data structures built using path hash accumulators[2] (see Section 4), and they result in recomputing the path hash accumulation(s) of the path(s) involved in these operations.

We next present and prove our first result about the performance and the authentication properties of path hash accumulators in the three-party authentication model described in Section 1.1.

**Lemma 1.** *Based on the path hash accumulator hashing scheme, there exists an authenticated data structure for path $p$ of length $n$ that supports query operations property(subpath), property(node), locate and subpath and update operations concatenate and split with the following performance:*

---

[2]Although the term path hash accumulator is defined as a hashing scheme (i.e., the authentication structure), we use the same term to also refer to the corresponding data structure (i.e., the binary tree); since the underlying DAG of the hashing scheme coincides with the binary tree, this introduces no confusion.

1. *query operations* property(subpath), property(node), locate *and* subpath *take each* $O(\log n)$ *time;*

2. *for every query operation the answer authentication information has size* $O(\log n)$;

3. *for every query operation the answer verification time is* $O(\log n)$;

4. *the total space used is* $O(n)$;

5. *for every update operation the update authentication information has size* $O(1)$;

6. *if $q$ is a path of length $m$, update operation* concatenate$(p,q)$ *takes* $O(\log(\max\{n,m\}))$ *and update operation* split *on $p$ takes* $O(\log n)$ *time.*

*Proof.* (1) First consider the query property($\bar{p}(v,u)$) on $p$. Let $\mathcal{A}(\bar{p})$ be the set of *allocation nodes* in $T(p)$ of subpath $\bar{p} = \bar{p}(v,u)$, defined as follows. For a tree node $w$, $w \in \mathcal{A}(\bar{p})$ if the leaves of the subtree defined by $w$ are all nodes of $\bar{p}$ but the same is not the case for $w$'s parent, if any. That is, $\mathcal{A}(\bar{p})$ is the minimal set of tree nodes defining subtrees that exactly cover $\bar{p}$ and no other nodes of $p$, i.e., subtrees whose leaves form a partition of $\bar{p}$ into subpaths, where each subpath consists of the leaves that correspond to one of the allocation nodes. So, each $w \in \mathcal{A}(\bar{p})$ corresponds to a subpath of path $\bar{p}$. Since $T(p)$ is a balanced binary tree, there are $O(\log n)$ allocation nodes for subpath $\bar{p}$ that can be found in $O(\log n)$ time by tracing the leaf-to-root tree paths in $T(p)$ from $v$ and $u$ up to the root $r$ of $T(p)$. Since, the path property $\mathcal{P}$ satisfies the concatenation criterion, we have that the path property $\mathcal{P}(\bar{p})$ can be computed by using the tree structure and by applying $O(\log n)$ times the concatenation function $\mathcal{F}$ of $\mathcal{P}$ on the path properties of the subpaths of $\bar{p}$ stored at the allocation nodes of $\bar{p}$. Thus, query property(subpath) can be answered in $O(\log n)$ time.

(2) Clearly, the answer given to the user is the property $\mathcal{P}(\bar{p})$. For any node $w$ of $T(p)$, let $(w_1, \ldots, w_k)$ be the node-to-root path connecting $w$ with the root $r$, with $w_1$ being $w$ and $w_k$ being a child of $r$. We define the *verification sequence* of $w$ to be the sequence $\mathcal{V}(w) = (s_1, s_2, \ldots, s_k)$, where, for $1 \leq j \leq k$,
$$s_j = (L(\bar{w}_j), \mathcal{P}(\bar{w}_j)), \tag{3}$$
and $\bar{w}_j$ is the sibling node of $w_j$, i.e., $s_j$ is the pair of the label of the sibling $\bar{w}_j$ of node $w_j$ and the path property of the path $p(\bar{w}_j)$ that corresponds to this sibling node $\bar{w}_j$. (Recall, property $\mathcal{P}(p(\bar{w}_j))$ is denoted simply as $\mathcal{P}(\bar{w}_j)$.) Let $z$ be the least common ancestor of $v$ and $u$ in $T(p)$. The answer authentication information except from the signed time-stamped digest consists of three parts (see Figure 1):

1. for each allocation node $w \in \mathcal{A}(\bar{p})$, the property $\mathcal{P}(w)$, if $w$ is not a leaf, or the property $\mathcal{N}(w)$ otherwise; these properties are given as a sequence $(\alpha_1, \ldots, \alpha_m)$, such that the set of leaf nodes of any allocation nodes with properties $\alpha_i$ and $\alpha_{i+1}$, $1 \leq i \leq m-1$, forms a subpath of $\bar{p}$;

2. for each allocation node $w \in \mathcal{A}(\bar{p})$, the labels of its children, if they exist, or the properties $\mathcal{N}(pred(w))$ and $\mathcal{N}(succ(w))$ otherwise; and

3. the labels and the corresponding path properties of the siblings of the nodes in the paths from the left most and right most allocation nodes up to the least common ancestor $z$, if these siblings are not allocation nodes themselves, and the verification sequence of $z$.

Given that $T(p)$ is balanced, thus, for every $\bar{p}$, the set $\mathcal{A}(\bar{p})$ of allocation nodes of $\bar{p}$ has size $O(\log n)$, and that a path property has constant size, the answer authentication information has size $O(\log n)$.

(3) To accept the answer, the user first recomputes $\mathcal{P}(\bar{p})$, by repeatedly applying the concatenation function $\mathcal{F}$ on sequence $(\alpha_1, ..., \alpha_m)$. If $\mathcal{P}(\bar{p})$ is not verified, the answer is rejected. Otherwise, the verification process is completed by the computation and verification of the signed path hash accumulation. Observe that the user has all the necessary information needed for this procedure. The computation of the digest corresponds to tracing two leaf-to-root paths in $T(p)$ and at each node of the paths computing a hash label and possibly applying the concatenation function $\mathcal{F}$. The answer authentication information can be given in such a way so that this sequence of computations is well-defined for the user; e.g., $O(\log n)$ bits can be used to denote the left-right relation of siblings in $T(p)$. Clearly, since computing the path hash accumulation corresponds to tracing two paths of length $O(\log n)$, where at each node a constant amount of work is performed or, equivalently, to executing an amount of computations proportional to the answer authentication information which has size $O(\log n)$, the answer verification time is $O(\log n)$.

(1) − (3) (*Other query operations*) Considering the other three query operations of the path hash accumulator, we note the following. For a property($v$) query, we proceed as above and the property($\bar{p}(v, u)$) case: observe that property($v$) corresponds to property($\bar{p}(v, v)$). For a locate($p,\sigma,q$) query, we locate the target node $v$ by performing a top-down search in $T(p)$ starting from the root: at a node $u$ with children $w_1$ and $w_2$, the path selection function $\sigma$ is used to select either the path that corresponds to $w_1$ or the path that corresponds to $w_2$. Then, the answer is the located node $v$ and the proof is the proof that corresponds to a property($\bar{p}(v, v)$) query. For a subpath($p,\alpha,q$) query, a similar top-down tree search is performed using the path advance function $\alpha$ to first compute the target subpath $\bar{p}(v, u)$; the proof is constructed by considering the corresponding allocation nodes. That is, the proof is the proof that corresponds to a property($\bar{p}(v, u)$) query. Thus, all these queries can be answered in $O(\log n)$ time, where the answer authentication information is of size $O(\log n)$ and the answer verification time is $O(\log n)$.

(4) Since a path property has constant size, the hash path accumulator occupies $O(n)$ space.

(5) Since the signed digest of the data set representing path $p$ is simply a hash value and, after any update operation on the path hash accumulator, 1 (concatenate) or 2 (split) digests are signed by the source and sent to the responder, the update authentication information has constant size.

(6) Besides, the update operations can be implemented in logarithmic time using the following primitive update operations on trees (see, e.g., [55]): create_root (given two trees, create a root that merges them into one), delete_root (delete the root of a tree to create two new trees) and rotate (perform a left or right rotation at a tree node). Observe that for all these operations the involved path properties can be computed or accordingly updated in constant time by applying the concatenation function $\mathcal{F}$. In particular, operation concatenate($p, q$) involves performing a create_root operation, computing the new hash path accumulation for $p\|q$, and then rebalancing the resulted new tree and accordingly updating the involved path properties and hash labels in this tree, through rotations in $O(\log(\max\{n, m\}))$ total time. Operation split involves performing the necessary rotations so that a delete_root operation creates the desired target trees, then rebalancing the resulted new trees and accordingly updating the involved path properties and hash labels in these trees, through the necessary "reverse" rotations in $O(\log n)$ total time. Overall, for both operations, updating the path hash accumulations corresponds to tracing at most two leaf-to-root paths and performing rotations and updating the hash labels at the visited nodes, thus resulting in logarithmic time complexities (see [62] for more details on these two path operations).

(*Security*) Considering the security provided by the hash path accumulator scheme, we note that we achieve the desired security results by reducing any attack from the responder against the user to a collision on the cryptographic hash function $h$ or a successful attack against the signature scheme in use by the source and the user. For every query, the answer authentication information includes the set of properties stored at the set of allocation nodes $\mathcal{A}(\bar{p})$ of a subpath $\bar{p}$ of $p$; these properties are checked against the query answer by the verification algorithm. Any attack against the validity of the answer corresponds to forging set $\mathcal{A}(\bar{p})$, i.e., providing the properties stored at a different set $\mathcal{A}'(\bar{p})$ of allocation nodes, which set of properties, of course, may possibly include one or more properties that do not correspond to any subpath of $p$, or one or more totally fake properties. A successful such attack corresponds to providing an incorrect path property that is not defined by the correct path hash accumulator, or including at least one node $w \notin \mathcal{A}(\bar{p})$ in $\mathcal{A}'(\bar{p})$, or omitting at least one node $w \in \mathcal{A}(\bar{p})$ from $\mathcal{A}'(\bar{p})$. By the path hash accumulator hashing scheme that takes into account the path properties as well as the left-right relation of siblings in $T(p)$ and the successor-predecessor relation of path nodes, and since for every subpath $\bar{p}$ we have that $head(\bar{p}) \in \mathcal{P}(\bar{p})$ and $tail(\bar{p}) \in \mathcal{P}(\bar{p})$, we have that all of the three above attack cases can be successfully detected by the user. It follows that in order to launch an attack the responder must forge the path hash accumulator hashing scheme, and in this case, the attack is successful either if the verified signed digest is not authentic, which corresponds to a successful attack against the security of the signature scheme, or if the modified hashing scheme produces at least one hash label that equals a hash value of the original, correct, hashing scheme, which corresponds to a successful attack against the collision-resistance of the cryptographic hash function. Therefore, if the set of allocation nodes that corresponds to the answer given to the user is not the correct one, in order for the user to accept the incorrect answer as correct, either a forgery against the signature scheme or at least one collision on $h$ must be computed by the responder. By the fact that our authentication scheme uses a secure signature scheme and a collision-resistant hash function, having a security parameter $k$ that equals the security parameter of the signature scheme and hash function in use, we have that any of these two events occurs with only negligible in $k$ probability. Therefore, our scheme satisfies the security requirement of Definition 2.4. □

The path hash accumulator can be viewed as a generalization of the Merkle's hash tree. That is, it provides a tree-based authentication scheme capable in authenticating more sophisticated queries than membership queries and also, as we will see in the next two sections, a general framework for building more complex authenticated data structures. We end this section with two useful remarks concerning Lemma 1.

*Remark* 3.1. Path property $\mathcal{P}(u)$ of subpath $p(u)$ is a sequence of constant size of path attributes. In Equation (2), $\mathcal{P}(u)$ participates in the hashing operation as the concatenation of a well-defined binary representation of the corresponding path attributes. That is, if $\mathcal{P}(u)$ consists of path attributes $a_1, \ldots, a_k$ that are related to path $p(u)$, then, by choosing to implement the hash of a sequence of values as the hash of their concatenation (see Section 2.2), Equation (2) reads $L(u) = h(a_1 \| \ldots \| a_k \| L(z_1) \| L(z_2))$. Similarly, in Equation (1), any node property $\mathcal{N}(w)$ participates in the hashing operation as the concatenation of a well-defined binary representation of the corresponding node attributes of node $w$.

*Remark* 3.2. For some specific applications, the path hash accumulator can be defined using a slightly different hashing scheme. In particular, we can hash the path property stored at a node of the tree before this is included in the label computation through hashing. That is, in the definition of the hashing scheme, we can replace the hash operation in Equation (2) with the operation

$$L(u) = h(h(\mathcal{P}(u)), L(z_1), L(z_2)). \tag{4}$$

This hashing scheme is more suitable in terms of performance in cases where the path property does not have constant size, but instead, it has size that is linear on the size of the path. We will encounter such a case for a specific type of queries in Section 4. We define the hash $h(\mathcal{P}(u))$ of the path property of subpath $p(u)$ to be (as in one of the examples in Section 2.2) the hash of the concatenation of the hashes of a well-defined binary representation of the path attributes in $\mathcal{P}$. That is, if $a_1, \ldots, a_k$, where $k$ is some constant, are the path attributes in $\mathcal{P}(u)$, then the hash of path property $\mathcal{P}(u)$ is $h(\mathcal{P}(u)) = h(a_1, \ldots, a_k) \triangleq h(h(a_1), \ldots, h(a_k))$. We note that the only changes that this modification of the hashing scheme brings to the results of Lemma 1 are: (*i*) the answer authentication information is now slightly different, but still of logarithmic size, namely, Equation 3 new reads

$$s_j = (L(\bar{u}_j), h(\mathcal{P}(\bar{u}_j))), \tag{5}$$

and (*ii*) according to the basic construction above, in the case where the path property has size that is proportional to the size of the path, the storage of the path hash accumulator becomes $O(n \log n)$. However, by introducing a special type of pointers into the data structure, a technique known with the name *threading*, we can actually reduce the storage needs of the data structure back to $O(n)$.

## 4 Authenticated Graph Searching

In this section, we consider authenticated data structures for graph searching problems. We wish to authenticate search queries on graphs, like queries that ask for a path connecting two nodes in a graph (if any), or for some information associated with this path, e.g., the size of the connecting path, as well as queries that ask structural information about the graph, e.g., queries about the connectivity between two nodes. Such data structures have applications to the authentication of network management systems.

Given the path hash accumulator authentication scheme described in the previous section, we follow a bottom-up approach in presenting our new authenticated data structures. We first develop a generic authenticated data structure for path property queries in a *forest*, that is, a collection of trees. The forest is dynamic, evolving through update operations that create, destroy, merge or separate trees. Trees store data items and querying information about these data items can be expressed by path property queries for paths in the trees of the forest for some path property that satisfies the concatenation criterion. Path hash accumulators are used as the primitive building blocks of this authenticated data structure. Furthermore, this new authentication structure for answering path properties queries on forests is then used to support more sophisticated queries on forests by appropriately defining the path property in use. Finally, we consider general graphs and use our authentication structures for forests to authenticate searching queries on these graphs.

### 4.1 Hierarchy of Paths

We start the construction of our authentication schemes by extending the use of path hash accumulators in *collections* of paths. In particular, we use the path hash accumulator authentication scheme over a dynamic collection $\Pi$ of paths that is maintained through the update operations split and concatenate defined over paths. At a high-level point of view, $\Pi$ is organized by means of a rooted tree $\mathcal{T}$ of paths, meaning that each node of $\mathcal{T}$ corresponds to a path in $\Pi$. Neighboring paths in $\mathcal{T}$ are generally interconnected and share information. This is achieved by the definition of suitable node attributes and properties.

A tree of paths $\mathcal{T}$ is considered to be directed; the direction of an edge is from a child to a parent. Let $\mu$ be a node of $\mathcal{T}$, let $\mu_1, \ldots, \mu_k$ be its children in $\mathcal{T}$ and let $p$ be the path that corresponds to node $\mu$. A node attribute $N(v)$ of a node $v$ of $p$ is extended so that it depends not only on $v$ but possibly also on some path properties of the paths $p_1, \ldots, p_k$ that correspond to nodes $\mu_1, \ldots, \mu_k$ of $\mathcal{T}$. We say that path $p$ is the *parent path* of paths $p_1, \ldots, p_k$ and these paths are the *children paths* of $p$. This extension of the semantics of a node attribute, i.e., allowing node attributes of a node of a path $p$ to be related to the path properties of $p$'s children paths, eventually allows the path property $\mathcal{P}(p)$ of path $p$ that correspond to node $\mu$ to include information about paths in the subtree of $\mathcal{T}$ rooted at node $\mu$. As always, we consider path properties that satisfy the concatenation criterion.

In general, the idea above can be further extended by using a directed acyclic graph (instead a tree) as the high level graph for the organization of a path collection $\Pi$. In fact, using such a graph, we introduce a *hierarchy* over paths in $\Pi$, where, accordingly, path properties are extended to include information (expressed by path properties) about other paths subject to the hierarchy induced by the graph.

## 4.2 Path Properties in a Forest

We now develop an efficient and fully dynamic authenticated data structure that supports path property queries in a forest, where the forest is realized as a hierarchy of paths. The data structure has fast, update, query, and verification times.

Let $\mathcal{N}$ be any node property and $\mathcal{P}$ be any (corresponding) path property. We assume that $\mathcal{P}$ satisfies the concatenation criterion. Let $F$ be a forest, a collection of trees. Forest $F$ is associated with a data set by storing at each tree node $u$ some information as node attributes, or equivalently as node property $\mathcal{N}(u)$. Additionally, using the framework presented in Section 3, any path $p$ in a tree of $F$ is associated with some path property $\mathcal{P}(p)$.

We study the implementation of the authenticated query operation property$(u, v)$—return the path property $\mathcal{P}$ of the path from $u$ to $v$ in $F$, if such a path exists—while the following update operations over trees in $F$ are performed:

- destroyTree$(w)$—it destroys the tree with root $w$;

- newTree()—it creates a new tree in $F$ that consists of a new, single, node;

- link$(u, v)$—it merges two trees into one by adding an edge between the root $u$ of some tree to a leaf $v$ of another tree;

- cut$(u)$—it separates a tree to two new trees by removing the edge between non-root node $u$ and its parent.

Note that any tree can be assembled or disassembled using these operations.

Our data structure is based on dynamic trees [55] introduced by Sleator and Tarjan. Conceptually, a dynamic tree $T$ is a rooted tree whose edges are classified (according to some criteria) as being either *solid* or *dashed*, with the property that any internal node of $T$ has at most one child connected by a solid edge. This edge classification partitions the tree into *solid paths*, i.e., consecutive nodes connected with each other through solid edges, whereas these solid paths are connected with each other by dashed edges (see Figure 2(a)). Note that a solid path may consist of only a single tree node that is incident to no solid edge. Using the framework of Section 3, we view every solid path of a dynamic tree as a path, i.e., a sequence of nodes, directed towards the root of $T$, that is, the successor of any node that is not the tail of the path is its parent node.

Moreover, by definition, every non-leaf node $v$ of a dynamic tree $T$ has at most one child $u_0$ such that a solid edge connects them. Assume that $v$ has more children and consider all these children, say nodes $u_1, \ldots, u_k$ in $T$ (connected with $v$ through dashed edges). Using again the framework of Section 3, we define the *dashed path* $d(v)$ of node $v$ to be a path, i.e., a sequence of nodes of length $k$, such that there is a one-to-one correspondence between edges $(u_i, v)$ in $T$ and path nodes of $d(v)$. The ordering of the nodes in $d(v)$ is thus in accordance with the ordering of nodes $u_1, \ldots, u_k$, which, in turn, can be arbitrary.

Having at hand the solid and dash paths defined in a dynamic tree, we now consider the trees of the forest $F$ to be dynamic trees, which allows us to perform a transformation of trees into solid and dash paths. In particular, let $T_1, \ldots, T_m$ be the trees in $F$. We view all these trees as dynamic trees. Let $\Pi(T_i)$ be the collection of all solid and dashed paths defined for tree $T_i$ of $F$ as explained above. Using the concept of hierarchies of paths discussed in Section 4.1, we can associate $\Pi(T_i)$ with a directed tree $\mathcal{T}_i$ of paths. This is performed as follows:

- each path $p$ (solid or dashed) in $\Pi(T_i)$ corresponds to a vertex $\mu_p$ of $\mathcal{T}_i$;

- if $p$ is solid, for each node $v$ of $p$ that has only one child $u$ in $T_i$ such that $u$ is node of path $p'$ in $\Pi(T_i)$ and $p \neq p'$ (and $v$ is connected with $u$ through a dashed edge), the directed edge $(\mu_{p'}, \mu_p)$ is an edge of $\mathcal{T}_i$;

- if $p' = d(v)$ is dashed with length $k$, that is, $p'$ corresponds to the dashed edges of a node $v$ in $T_i$, let $p$ be the solid path that $v$ belongs to, let $u_1, \ldots, u_k$ be the corresponding children of $v$ in $T_i$, and let $p_1, \ldots, p_k$ be the solid paths containing these children; then, the directed edges $(\mu_{p'}, \mu_p)$ and $(\mu_{p_i}, \mu_{p'})$, $1 \leq i \leq k$, are edges of $\mathcal{T}_i$.

Finally, given the directed trees of paths $\mathcal{T}_i$, $i = 1, \ldots, m$, we add a new *root vertex* $\omega$ which is the parent of all the roots of trees $\mathcal{T}_i$, thus, obtaining a new tree $\mathcal{F}$ (representing the entire forest $F$).[3] All the newly added edges are directed towards $\omega$. We consider one last *root path* $\pi(\omega)$ that corresponds to the root vertex $\omega$. The nodes of this path correspond to trees $T_i$ of $F$, where any node ordering in $\pi(\omega)$ can be used. Our final graph is a tree of solid and dash paths rooted at this special root path $\pi(\omega)$.

Consider the collection $\Pi(F)$ of paths (solid, dash, root) associated with the nodes of tree $\mathcal{F}$. The children of the root path $\pi(\omega)$ are solid paths. The children of a solid path are either solid of dashed paths. The children of a dashed path are solid paths. Figure 2(b) shows such a tree $\mathcal{F}$ (corresponding to the forest $F$ of Figure 2(a)).

Using this tree of paths, we implement our data structure as follows. Each path (root, solid or dashed) is implemented through the path hash accumulator authentication scheme (Section 3), where the individual data structure that implements each path hash accumulator is chosen to be a biased binary tree [6].

---

[3]Root vertex $\omega$ is a fictitious node, used only as a means to define a special root path on top of the trees $\mathcal{T}_i$.

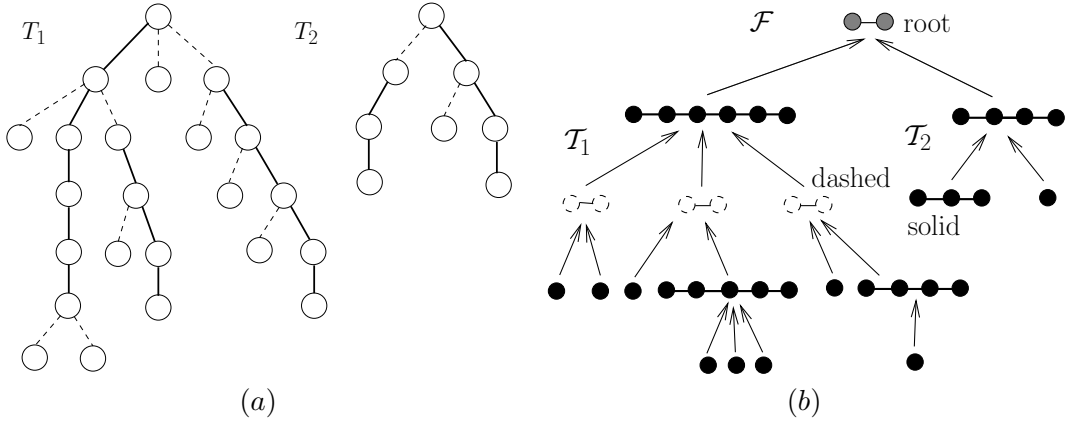

Figure 2: (a) The partition of trees into solid paths. (b) Trees of paths and final tree $\mathcal{F}$.

A path property $\mathcal{P}$ of our interest, i.e., a collection of path attributes, that satisfies the concatenation criterion is defined. By the implicit path interconnection, through the idea of setting a path property of a path to be a node attribute of another neighboring path, $\mathcal{P}(p)$ includes information about the children paths of $p$ and, in general, about all of its descendant paths in the hierarchy of paths induced in $\mathcal{F}$. We include path attributes in node properties, as follows. If $v$ is a node of path $p$ and $L(p')$ denotes the path hash accumulation of path $p'$, then:

1. if $p$ is root path or dashed path, then $L(p')$ and $tail(p')$ are included in $\mathcal{N}(v)$, where $p'$ is the child solid path of $p$ corresponding to node $v$;

2. if $p$ is solid path, then $L(p')$ and $tail(p')$ are included in $\mathcal{N}(v)$, if $v$ corresponds to a solid child path $p'$ of $p$, or $L(p')$ is included in $\mathcal{N}(v)$, if $v$ corresponds to a dashed child path $p'$ of $p$.

The above scheme of inclusions of path attributes and path properties of a path as a node attribute in the node property of a node of the parent path corresponds to connecting the individual hashing schemes of the path hash accumulators implementing paths in $\mathcal{F}$ and composing them into one hashing scheme $G$ for the entire data structure. This hashing scheme $G$ yields a digest for the entire data set that is stored in the forest $F$, namely, the path hash accumulation of the root path $\pi(\omega)$ of $\mathcal{F}$. Next, we present our first theorem, which fills in the details of the entire authenticated data structure that supports update and query operations on forest $F$, analyze its performance and prove its efficiency.

**Theorem 2.** *Let $F$ be a forest of $t$ trees with $n$ nodes. There exists a fully dynamic authenticated data structure that supports query operations **property** on paths in dynamic forest $F$ that evolves through update operations **destroyTree**, **newTree**, **link** and **cut**, having the following performance:*

1. *query operation **property** takes $O(\log n)$ time;*

2. *for query operation **property** the answer authentication information has size $O(\log n)$;*

3. *for query operation **property** the answer verification time is $O(\log n)$;*

4. *the total space used is $O(n)$;*

5. for every update operation the update authentication information has size $O(1)$;

6. update operations destroyTree and newTree take $O(\log t)$ time each; update operations link and cut take $O(\log n)$ time each.

*Proof.* (*Data Structure*) We first complete the description of the data structure. As we already have seen, the entire forest $F$ is represented as a collection of paths $\Pi(F)$, which is organized using the hierarchy induced by the tree $\mathcal{F}$ of paths in $\Pi(F)$. Recall that the solid paths in $\Pi(F)$ are defined by considering each tree $T_i$ of $F$ to be a dynamic tree and by using a partition of edges into solid and dashed. In any tree $T_i$, let $\mathsf{size}(v)$ denote the number of nodes in the subtree defined by $v$ and let $u$ the parent node of $v$. Edge $e = (v, u)$ is called *heavy* if $\mathsf{size}(v) > \mathsf{size}(u)/2$. The edge labeling of dynamic tree $T_i$ of $m_i$ nodes with root $w$, such that an edge is labeled solid only if it is heavy, has the following important property [55]: for any node $u$ of $T_i$ there are at most $\log m_i$ dashed edges on the path from $u$ to $w$. We use this edge labeling to partition each tree $T_i$ into solid paths.

Consider all the paths that correspond to the final tree $\mathcal{F}$, after the dashed paths and the root path have been added. Each path $p$ of $\mathcal{F}$ is represented using the path hash accumulator authentication scheme of Section 3, with only one exception. Following the ideas in [55], path $p$ is represented by a binary tree $T(p)$, but $T(p)$ is implemented as a biased binary tree $T(p)$ (see [6]), thus, it is not necessarily height balanced. Note that this fact does not affect the corresponding hashing scheme; the path hash accumulation $L(p)$ of $p$ is still well-defined.

In a biased binary tree, each leaf node is associated with a weight, each non-leaf node is associated with the sum of the weights of its children, (consequently) tree's root $r$ carries the sum $W$ of the weights of the leaves, and any node $v$ with weight $w(v)$ lies at depth $O(\log \frac{W}{w(v)})$. In our data structure and as in the data structure of [55], node weights are defined using function $\mathsf{size}()$, where we consider weight $w(v)$ of node $v$ to be an additional node or path attribute, depending on whether $v$ is a leaf node in $T(p)$ or not.[4] If $p$ is a path having no child path ($\mu_p$ is a leaf in $\mathcal{F}$), then $w(v) = \mathsf{size}(v)$. Otherwise ($\mu_p$ is not a leaf in $\mathcal{F}$), $w(v) = w(u_1) + w(u_2)$, if $v$ is internal node of $T(p)$ with children $u_1, u_2$. Otherwise, $v$ is a node of path $p$. If $p$ is solid, then $w(v) = w(u) + 1$, where $u$ is the root of $T(p')$ and $p' = d(v)$ is the dashed path of $v$, if such dashed path exists, or $w(u) = 0$, if $v$ does not have a dashed path. If $p$ is dashed, then $w(v) = w(u) + 1$, where $u$ is the root of $T(p')$ and $p'$ is the solid, child path of $p$ in $\mathcal{F}$ that corresponds to $v$. If $p$ is root path, then $w(v) = w(u) + 1$, where $u$ is the root of $T(p')$ and $p'$ is the child path of $p$ in $\mathcal{F}$ corresponding to $v$.

To complete the description of the authenticated data structure, we note that the hashing scheme of the entire data structure is defined through the individual path hash accumulator hashing schemes of the paths in $\mathcal{F}$. All these hashing schemes compose a hashing scheme for the entire forest $\mathcal{F}$ and the date structure as a whole, where the path hash accumulation of a path $p$ contributes to the computation of the path hash accumulation of $p$'s parent path in $\mathcal{F}$.

(*Efficiency*) Consider any tree $T_i$ in $F$, any two nodes $u$ and $v$ of $T_i$ and the path $p_{uv}$ in $T_i$ that connects $u$ and $v$. Our data structure represents $T_i$ implicitly through the path collection $\Pi(T_i)$, where each path in $\Pi(T_i)$ is implemented as a (biased) binary tree. For answering queries regarding properties of path $p_{uv}$, we first consider a *multipath* $\pi_{uv}$, that is, a path of paths in $\mathcal{T}_i$, that connects the paths in $\Pi(T_i)$ where $u$ and $v$ belong, and using this multipath, we then consider a (different of $p_{uv}$) path $P_{uv}$: the path that virtually connects $u$ and $v$ in $\mathcal{T}_i$ *through* the binary trees that implement the paths of $\Pi(T_i)$. Path $P_{uv}$ is a connecting path of nodes $u$ and $v$ in the hashing scheme of the entire data structure, when no edge directions are taken into consideration. These three type of paths $p_{uv}$, $\pi_{uv}$ and $P_{uv}$ are described in Figure 3.

---
[4]Related to the sum function and as any *aggregate* function (e.g., max, average), function $w(\cdot)$ satisfies the concatenation criterion; thus, for convenience (and not for authentication reasons) it can be treated as a path property.

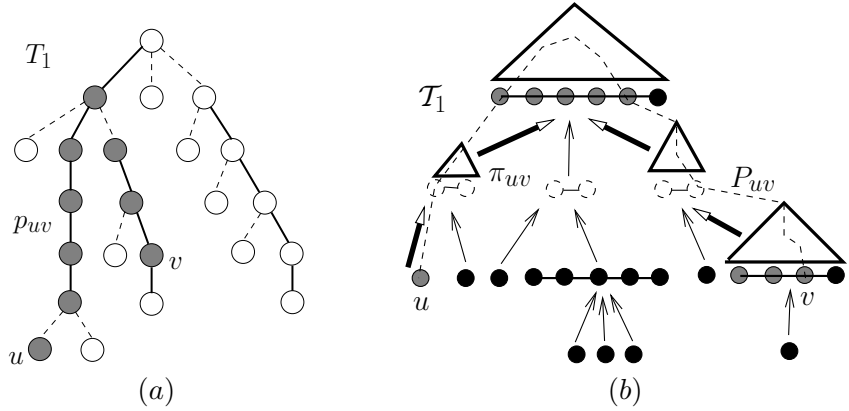

Figure 3: (a) Path $p_{uv}$ in tree $T_1$ (grey nodes), connecting nodes $u$ and $v$. (b) Multipath (path of paths) $\pi_{uv}$ in $\mathcal{T}_1$ (indicated by dark arrows), connecting the paths containing $u$ and $v$, and finally path $P_{uv}$ (indicated by dashed line) in the data structure, connecting nodes $u$ and $v$ through the (biased) binary trees implementing paths in $\Pi(\mathcal{T}_1)$. Triangles denote biased binary trees, not necessary height balanced. Path $P_{uv}$ is also a path in the hashing scheme $G$ of the entire data structure. Observe that $P_{uv}$ defines allocation nodes in binary trees implementing solid paths in $\mathcal{T}_1$, which allocation nodes correspond to subpaths of $p_{uv}$ and store path properties that constitute path property $\mathcal{P}(p_{uv})$. Through the biased trees, $P_{uv}$ has logarithmic length on the size of $T_1$.

Observe that path $P_{uv}$ passes through nodes of the binary trees implementing solid paths in $\mathcal{T}_i$ that have as children, tree nodes defining subtrees whose leaves are subpaths of $p_{uv}$, that is, nodes that are allocation nodes of subpaths of path $p_{uv}$. This is exactly what is needed: in authenticating a path property of $p_{uv}$, we will use the path properties stored at nodes (of the biased binary trees) related to path $P_{uv}$ that, given the fact that the path property in study satisfies the concatenation criterion, completely describe the path property of $p_{uv}$. Accordingly, the above considerations are also valid for paths in the data structure that connect nodes $u$ and $v$ of different trees in $F$: in this case, a multipath in $\mathcal{F}$ that passes through the root path $\pi(\omega)$ connecting the paths in $\Pi(F)$ of nodes $u$ and $v$ exists and the corresponding path $P_{uv}$, connecting $u$ and $v$ in the hashing scheme of the data structure, passes through the binary tree implementing the root path $\pi(\omega)$.

Using of the previously described biasing in our data structure, it holds that, when considered *through* the individual biased binary trees that implement paths in $\mathcal{F}$, any path $P_{uv}$ in tree $T_i$ of size $m_i$ has length $O(\log m_i)$ and any leaf-to-root path in the data structure representing $\mathcal{F}$ has length $O(\log n)$. The proof is based exactly on the analysis of the properties of dynamic trees studied in [55], the idea being that through the solid-dashed path partition and the implementation of paths as biased binary trees, any two nodes in a (possibly highly unbalanced) tree are connected—through shortcuts—with paths of total logarithmic size in the tree size, effectively in a new, combined, balanced tree structure.

$(1) - (3)$ Consider query property$(u, v)$. Although this query is defined for nodes in forest $F$, to cover the most general case, we do not require that query nodes $u$ and $v$ necessarily exist in $F$. So, we first determine whether nodes $u$ and $v$ are in forest $F$ using any authenticated data structure that supports membership queries (e.g., [29]) in $O(\log n)$ time with authenticated responses of size $O(\log n)$. If one of the two nodes are not in $F$, a negative answer is given, along with a proof that verifies the negative membership. If $u$ and $v$ are in $F$, the path property query is performed by accessing three multipaths in $\mathcal{F}$ (see Figure 4).

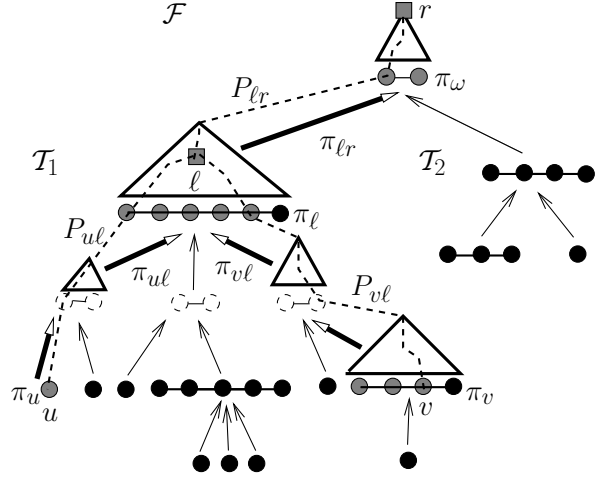

Figure 4: The answer and proof for property$(u,v)$ query are computed by visiting paths $\pi_{u\ell}$, $\pi_{v\ell}$ and $\pi_{\ell r}$ in the data structure and by accessing information stored at nodes related to these paths.

In particular, assume first that $u$ and $v$ belong to the same tree $T_i$ of size $m_i$ in $F$, i.e., there exists a connecting path $p_{uv}$ in $T_i$. Let $\pi_u$ and $\pi_v$ be the paths in $\mathcal{F}$ that contain $u$ and $v$ and let $\pi_\ell$ be the least common ancestor of $\pi_u$ and $\pi_v$ in $\mathcal{F}$ ($\pi_l$ may overlap with $\pi_u$ and/or $\pi_v$). Let $\pi_{u\ell}$, $\pi_{v\ell}$ be the multipaths from $\pi_u$, $\pi_v$ to $\pi_\ell$, and let $\pi_{\ell r}$ be the multipath from $\pi_l$ to the root path $\pi(\omega)$ of $\mathcal{F}$. Multipaths $\pi_{u\ell}$, $\pi_{v\ell}$ and $\pi_{\ell r}$, when considered *through* the biased binary trees used to implement paths in $\mathcal{F}$, define paths $P_{u\ell}$, $P_{v\ell}$ and $P_{\ell r}$, respectively, in the data structure. Specifically, paths $P_{u\ell}$ and $P_{v\ell}$ are connected at node $\ell$ of the binary tree implementing path $\pi_\ell$ (node $\ell$ is the least common ancestor of $u$ and $v$ in the data structure). The answer $A(u,v)$ given to the user is computed by following paths $\pi_{u\ell}$ and $\pi_{v\ell}$, finding, at each traversed path hash accumulator that implements a solid path in $\mathcal{F}$, the allocation nodes whose subtrees correspond to subpaths of $p_{uv}$ and, finally, reporting the path properties stored at these allocation nodes. Similarly, the proof given to the user is collected by providing path property proofs (i.e., collection of appropriate hash values and path properties as described in the proof of Lemma 1 in Section 3) $proof(\pi_{u\ell})$, $proof(\pi_{v\ell})$, each corresponding to a traversed multipath to $\pi_\ell$ and consisting of subproofs, one for each path hash accumulator that is visited. To compute the proof, following path $P_{\ell r}$ up to the root $r$ of the binary tree implementing path $\pi(\omega)$, we compute the *multipath verification sequence* $\mathcal{V}(\ell)$, a collection of hash values and path properties (similar to the verification sequence in the proof of Lemma 1) that allows the user to recalculate the signed hash value (hash path accumulation of $\pi(\omega)$) given $A(u,v)$, $proof(\pi_{u\ell})$ and $proof(\pi_{v\ell})$. By the biased scheme used over $\mathcal{F}$, the set of allocation nodes of path $p_{uv}$ has size $O(\log m_i)$ and also paths $P_{u\ell}$, $P_{v\ell}$ and $P_{\ell r}$ have each $O(\log n)$ size. Since path properties have constant size and for each node of these paths visited in the data structure a constant amount of information is included in the proof and a constant amount of work is performed, the answer $A(u,v)$ has $O(\log n)$ size, the proof $(\mathcal{V}(\ell), proof(\pi_{u\ell}), proof(\pi_{v\ell}))$ has also size $O(\log n)$ and path properties queries are answered in $O(\log n)$ time. Accordingly, the verification time is also $O(\log n)$.

Similarly, in the special case[5] where no path connecting $u$ and $v$ exists in $F$ ($u$ and $v$ belong in different trees), a negative answer is given to the user, indicating that no path connecting the query nodes exists in $F$) and the previous approach is used to provide proof of this fact: proofs corresponding to paths $P_{u\ell}$, $P_{v\ell}$ and $P_{\ell r}$ verify the nonexistence of path $p_{uv}$, since the least common ancestor $\ell$ of $u$ and $v$ in the data structure is node of the binary tree implementing root path $\pi(\omega)$.

(4) Since a path property has constant size, the hash path accumulator occupies $O(n)$ space.

(5) Since the signed digest of the entire data structure is simply a hash value, the path hash accumulation of the root path in $\mathcal{F}$, the update authentication information has constant size.

(6) All update operations correspond to accessing and modifying multipaths through the primitive path operations split and concatenate. In particular, operations link and cut can be implemented in $O(\log n)$ time by modifying only $O(\log n)$ path hash accumulators and by examining, modifying and restructuring only $O(\log n)$ nodes in total. Restructuring means connecting a node to new children. Our scheme works by, every time a node $v$ is restructured, recalculating $L(v)$, which can be done in $O(1)$ time, since the hash values and the path properties of the children, parent or neighbors of $v$ are known, and because the path property in consideration satisfies the concatenation criterion. Consequently, our update operations link and cut can be performed in $O(\log n)$ time. Finally, the update operations destroyTree and newTree each involves modifying the path hash accumulator that corresponds to the root path in $\mathcal{F}$, which has $O(t)$ size. Therefore these two update operations take $O(\log t)$ time.

(*Security*) Hashing scheme $G$ is based on the path hash accumulator. By allowing neighboring (in $\mathcal{F}$) paths to share information (properties) we achieve the desired security results based on the security of path hash accumulator authentication scheme: any attack to our data structure can be reduced to an attack on the security of the path hash accumulator, thus, in turn to either a collision on the cryptographic hash function $h$ or an attack against the signature scheme in use. In particular, the security arguments in the proof of Lemma 1 are generalized to the authenticated data structure for path properties in forests in trees, as follows. Any successful attack by the responder amounts to constructing a set of allocation nodes corresponding to a set of path properties that is distinct to the correct set of path properties defined by the answer of the query and the hashing scheme. Therefore, the responder must forge some path properties of at least one (solid, dashed or root) path in $\mathcal{F}$. Through the interconnection of the path hash accumulators in our data structure to a single hashing scheme, this corresponds to forging the hashing scheme of at least one hash path accumulator or to directly forging the signature on the data set digest. This yields the desired reduction to the security of the path hash accumulator which holds with all but negligible in the security parameter probability from Lemma 1. □

### 4.3 Path, Connectivity and Type Queries on Forests

Theorem 2 supports the basis for an authenticated data structure that efficiently answers and authenticate the following queries on a dynamic forest $F$:

- path($u$, $v$): reports the path, if any, between nodes $u$ and $v$ in $F$;

---

[5]This case is included in the proof only for completeness, since, by definition, a path property query returns the path property of existing in $F$ paths. Our used approach is similar with the one used in Section 4.3 for the authentication of higher-level path and connectivity queries: the nonexistence of a path in $F$ is authenticated by the existence of a path in $\mathcal{F}$ passing through the root path $\pi(\omega)$.

- pathLength($u$, $v$): reports the length of the path, if any, between nodes $u$ and $v$ in $F$;

- areConnected($u$, $v$): reports whether there is a path between nodes $u$ and $v$ in $F$ (i.e., whether $u$ and $v$ are nodes of the same tree);

- type($u$, $v$): reports whether there is a node of a given type in the path, if any, connecting nodes $u$ and $v$ in $F$; here, we assume that the type of a node is a well-defined notion that can be checked in constant time.

Our results are obtained by appropriately defining a path property that expresses each one of the above queries. In other words, each new query is answered as a path property query, as in Section 4.2, for a specially defined path property, which, of course, satisfies the concatenation criterion. Each specific query corresponds to a specially chosen individual path attribute that is included in the path property in use. The next theorem presents the details of these results.

**Theorem 3.** *Let $F$ be a forest of $t$ trees with $n$ nodes. There exists a fully dynamic authenticated data structure that supports query operations* path, pathLength, areConnected *and* type *on paths in dynamic forest $F$ that evolves through update operations* destroyTree, newTree, link *and* cut *having the following performance, where $k$ is the length of the path returned by operation* path:

1. *query operations* pathLength, type *and* areConnected

    - *each takes $O(\log n)$ time;*
    - *each has answer authentication information of size $O(\log n)$;*
    - *each has $O(\log n)$ answer verification time;*

2. *query operation* path

    - *takes $O(\log n + k)$ time;*
    - *has responder-to-user communication cost [6] of size $O(\log n + k)$, where the answer has size $O(k)$ and the answer authentication information has size $O(\log n)$;*
    - *has $O(\log n + k)$ answer verification time;*

3. *the total space used is $O(n)$;*

4. *for every update operation the update authentication information is $O(1)$;*

5. *update operations* destroyTree *and* newTree *take $O(\log t)$ time each; update operations* link *and* cut *take $O(\log n)$ time each.*

*Proof.* (*Data Structure*) It is essentially the same data structure as the one of Theorem 2, except from two differences, one with respect to the support of query path and one with respect to the exact form of the hashing scheme of the data structure, designed to simultaneously support all queries operations. In what follows, we explain in detail these differences, however noting that an alternative implementation of our data structure could be as follows: the data structure simply consists of the aggregation of four different hashing schemes, one for each supported query operation; that is, each query operation corresponds to a property query for a specially defined path property.

---

[6]Since the answer size in this case is not constant, we analyze the responder-to-user communication cost and accordingly distinguish the costs of the answer and the answer authentication information.

The path property $\mathcal{P}$ in use—except, by definition, from (the $id$s of) the head and the tail nodes of the corresponding path—includes four specially chosen path attributes $a_1$, $a_2$, $a_3$ and $a_4$ one for each of the supported query operations, respectively, path, pathLength, areConnected and type. In this way, path property $\mathcal{P}$ contains information for all supported queries and, more importantly, it has size that is *linear* on the size of the corresponding path.

Because of this, we use the implementation of the hashing operation of Equation 4 in defining the hashing scheme of our data structure. In our implementation, the hashing operation over a path property is accordingly designed (see Section 2.1 and Remark 3.2) so that more efficiency is achieved. In particular, the path property $\mathcal{P}$ is viewed as a sequence of path attributes $a_1$, $a_2$, $a_3$, $a_4$ and, accordingly, its hash value is computed as

$$h(\mathcal{P}) = h(h(a_1), h(a_2), h(a_3), h(a_4)). \tag{6}$$

That is, the hash of a path property is the hash of the concatenation of the hash values of the attributes that it consists of. The time complexity in computing this hash value is linear on the size of the path.

Overall, the above choice for implementing the hashing operation used in the definition of the hashing scheme has two important consequences. First, the answer authentication information is still logarithmic in the size of the forest (regardless the fact that the path property has size proportional to the path). Second, query operations can be answered separately, where path attributes are treated not as a whole (and not as a single path property query as in Theorem 2), but as individual pieces of information, namely individual path attributes. That is, attribute $a_1$ of a path, defined in the straightforward way as the ids of the nodes contained in the path, can be authenticated without authenticating or revealing any of the other three attributes. In essence, using this hashing scheme, a path property query can be answered as in Theorem 2 but *only with respect to* a specific path attribute.

Moreover, regarding the concatenation function $\mathcal{F}$ of property $\mathcal{P}$, it is simply defined as a per-attribute application of function $\mathcal{F}$. That is, if $p = p' \| p''$, $\mathcal{P}(p') = (a'_1, a'_2, a'_3, a'_4)$ and $\mathcal{P}(p'') = (a''_1, a''_2, a''_3, a''_4)$, then we have that

$$\mathcal{P}(p) = \mathcal{F}((a'_1, a'_2, a'_3, a'_4), (a''_1, a''_2, a''_3, a''_4)) = (\mathcal{F}(a'_1, a''_1), \mathcal{F}(a'_2, a''_2), \mathcal{F}(a'_3, a''_3), \mathcal{F}(a'_4, a''_4)).$$

Additionally, since the path property is not of constant size, the concatenation function $\mathcal{F}$ operates in time proportional to the path property.

In what follows, we define the path attributes, discuss their concatenation function $\mathcal{F}$ and complexity issues that are relevant to our authenticated data structure.

(1) Clearly, operation path corresponds to the path attribute $a_1$ containing the ids of the nodes that a path consists of, and the concatenation function is exactly the concatenation operator over sequences of path nodes.

For query operation pathLength, we define the path attribute $a_2$ of a path to be simply the size of the path and the concatenation function is the addition function.

Query operation areConnected corresponds to the existence of a node of the root path $\pi(\omega)$ in the path $P_{uv}$ connecting nodes $u$ and $v$ in the data structure representing forest $F$ (or tree $\mathcal{F}$). Note that path $P_{uv}$ always exists. That is, the answer to the query is negative if such a node exists in $P_{uv}$ and positive if no node of the root path $\pi(\omega)$ exists in $P_{uv}$. This property is expressed by assigning a unique $id$ value to every path in the tree of paths $\mathcal{F}$. Thus, attribute $a_3$ of a path $p$ in the tree of paths $\mathcal{F}$ and accordingly in our data structure is defined to take on one of the following two values: root-path-true, if $p$ is a subpath of $\pi(\omega)$, or root-path-false. Concatenation function $\mathcal{F}$ operates in constant time as the OR boolean function on pairs of these values of $a_3$.

A similar idea is applied for query operation type. A node attribute corresponding to path attribute $a_1$ (or path attribute $a_4$ of path of size 1) takes on two values: either type-true or type-false, depending on whether on not the corresponding node is of the type of interest (that the query operation asks about). Again, the concatenation function operates as the boolean function OR.

The complexity for all these three last query operations is similar to the complexity of the path property query operations of Theorem 2. Although, path attribute $a_1$ has size that is proportional to the size of the corresponding path, because of the hashing scheme, which hashes individually the path attributes in producing the hash of a path property (Equation 6), no path attribute $a_1$ is included in the answer or in the answer authentication information for any of the query operations pathLength, areConnected or type.

(2) Query operation path is answered by first performing a query areConnected. If there is a path between nodes $u$ and $v$, it can be found by answering a path property query with respect to path attribute $a_1$, where attribute $a_1$ of path $p$ includes all the (*ids* of the) nodes of path $p$, that is, the path itself. To this end, we need a slightly different definition for the path attribute, namely, a path attribute can be of any size (not necessarily constant). Since $|a_1| = O(|p|)$, the introduced complexity is $O(\log n + k)$, where $k$ is the length of the path from $u$ to $v$. In particular, both the query time and the answer verification time are $O(\log n + k)$, since the answer itself is of size $O(k)$ and both computations need to spend time proportional to the answer (to compute and process the answer respectively). The logarithmic term is due to the complexity that is carried from Theorem 2. On the other hand, the answer authentication information is still logarithmic, since our hashing scheme applies an extra hashing operation over the path property, before hashing over the sibling hash labels (see Equation 4).

(3) Although, according to the construction in Section 3 the path hash accumulator for a path property of linear size has $O(n \log n)$ storage, using *threads* in our data structure, we can reduce the storage needs to $O(n)$. The idea is to store path attribute $a_1$ of path $p$, the one corresponding to the subtree defined by internal node $u$, at the leaves of this subtree (that is, at the path $p$ itself), rather than storing it at node $u$. Then, we add a special pointer from node $u$ to $head(p)$ and a special pointer from $tail(p)$ up to node $u$ and pointers for every node in solid path towards its successor node in the path (if any). These pointers, called threads, can be used to traverse path $p$ in $O(|p|)$ time and compute (retrieve) the path attribute $a_1$. Thus, the total storage can still be kept linear, even with the presence of a path property of non constant, but linear, size.

(4), (5) & (*Security*) They follow directly from Theorem 2. □

We next show how this result can be extended to give authenticated schemes for more advanced graph queries. These results can be applied to the authentication of network management systems.

## 4.4 Path and Connectivity Queries on Graphs

We now move our attention to graphs rather than forests. Suppose we want to authenticate path and connectivity queries on a general graph $G$. That is, as before, we want to authenticate the answers to the following queries on $G$:

- path($u$, $v$): report the path, if any, between nodes $u$ and $v$ in $G$;

- areConnected($u$, $v$): report whether there is a path between nodes $u$ and $v$ in $G$ (i.e., whether $u$ and $v$ are nodes of the connected component in $G$).

We can immediately apply Theorem 3 to design an authenticated data structure for path and connectivity queries in a graph $G$ that evolves through vertex and edge insertions. In particular, graph $G$ is maintained through update operations:

- makeVertex($v$): create a new vertex $v$ in $G$;
- insertEdge($u, v, e$): add edge $e$ between vertices $u$ and $v$ in $G$.

The new data structure has similar performance bounds with the one in Theorem 3. The main idea in our data structure is to maintain a spanning forest $F$ of the graph $G$. We note that for embedded planar graphs our data structure can actually be extended to also support deletions of vertices and edges, through new update operations:

- destroyEdge($e$): destroy edge $e$ in $G$; and
- destroyVertex($u$): destroy isolated vertex $u$ in $G$.

The idea is to use techniques similar to the data structure described in [23].

**Theorem 4.** *Let $G$ be a general graph with $t$ connected components and $n$ nodes. There exists a semi-dynamic authenticated data structure that supports query operations path, and areConnected on pairs of nodes in graph $G$ that evolves through update operations makeVertex and insertEdge having the following performance, where $k$ is the length of the path returned by operation path:*

1. *query operation areConnected takes $O(\log n)$ time and query operation path takes $O(\log n + k)$ time;*

2. *for query operation areConnected the answer authentication information has size $O(\log n)$; for query operation path the responder-to-user communication cost has size $O(\log n + k)$, where the answer has size $O(k)$ and the answer authentication information has size $O(\log n)$;*

3. *for query operation areConnected the answer verification time is $O(\log n)$; for query operation path the answer verification time is $O(\log n + k)$;*

4. *the total space used is $O(n)$;*

5. *for every update operation the update authentication information is $O(1)$;*

6. *update operation makeVertex takes $O(\log t)$ time; update operation insertEdge takes $O(\log n)$ time;*

7. *if $G$ is an embedded planar graph, additional update operations destroyEdge and destroyVertex are supported in $O(\log n)$ and $O(1)$ time, respectively.*

*Proof.* (*Data Structure*) The idea is to use the data structure of Theorem 3 to maintain a *spanning forest* of graph $G$. That is, we maintain a forest $F$ that spans through the entire graph, meaning that all nodes in $G$ are nodes of $F$ as well and that each connected component of $G$ corresponds to a tree of $F$. The data structure of Theorem 3 allows us to authenticate answers to path and connectivity queries. Note the correctness of the data structure: if two nodes are connected in $G$, they are connected in $F$ as well, for they belong to the same connected component of $G$, thus to the same tree in $F$, and in this case, the path connecting them in this tree is obviously a path connecting them in $G$ as well.

Update operations are handled as follows. For general graphs, where only vertex and edge insertions are supported, each new vertex corresponds to a new connected component of $G$, thus, to simply a new tree in $F$ and a newTree update operation, whereas each new edge either corresponds to no action, when it connects nodes of the same tree (connected component in $G$) in $F$, or it corresponds to a link operation, when it connects nodes of different trees (connected components in $G$) in $F$. Testing whether or not an edge connects nodes of the same tree in $F$ can be performed by assigning unique $id$s to all trees in $F$ and checking whether or not the two nodes belong in trees with the same $id$. The last operation can be done by simply storing at each node of $F$ the corresponding tree $id$ (or even by accordingly defining an additional path attribute).

Update operations for embedded planar graphs, which include not only edge and vertex insertions but also deletions, are a bit trickier to handle. First, any insertion of an edge connecting nodes of the same tree needs to be stored. After any deletion of an edge or a vertex, the set of stored edges that are not edges in $F$ and, thus, not explicitly stored in the data structure representing $F$, is processed to decide whether or not this deletion results in a connected component destruction in $G$. Using the data structure described in [23], one can decide on whether a new connected component is created or, instead, the connected component stays the same but with a different spanning tree this time. Both this decision and the update of the component can be done in logarithmic time on the size of the graph, using the fact that edges admit efficient representation because of the planarity property of the graph $G$. The use the data structure in [23] is an orthogonal issue in our data structure, meaning that it is not connected with the operation of the authenticated data structure, but rather, it supports the maintenance of the forest $F$. Once forest $F$ is updated—always through the update operations that the data structure supports—the hashing scheme is accordingly updated and the new digest is computed.

$(1) - (7)$ & ($Security$) They follow immediately from Theorem 3. $\square$

In the next two subsections we use the results of Theorems 2 and 3 for connectivity queries for general graphs that evolve through edge and vertex insertions. We use known techniques (data structures) that support these type of queries for regular (non-authenticated) data structures and apply our authentication framework of Section 3, which is based on paths and their properties, on these data structures by appropriately authenticating path properties that are related to the connectivity queries that we study. In other words, here, we have the first applications of our authentication framework of the path hash accumulator to the authentication of queries that are not directly related to paths.

### 4.5 Biconnectivity Queries on Graphs

As before, let $G$ be a general graph that is maintained through update operations makeVertex and insertEdge. We are interested in authenticating the query operation:

- areBiconnected$(u, v)$: determine whether $u$ and $v$ are in the same biconnected component of $G$,

which we call a *biconnectivity* query. Theorems 2 and 3 can be used to support an authenticated data structure that answers biconnectivity queries.

**Theorem 5.** *Let $G$ be a general graph with $t$ connected components and $n$ nodes. There exists a semi-dynamic authenticated data structure that supports query operation areBiconnected on pairs of nodes in graph $G$ that evolves through update operations makeVertex and insertEdge having the following performance:*

1. *query operation areBiconnected takes $O(\log n)$ time; for this query operation, the answer authentication information has size $O(\log n)$ and the answer verification time is $O(\log n)$;*

2. *the total space used is $O(n)$;*

3. *for every update operation the update authentication information is $O(1)$;*

4. *update operation makeVertex takes $O(\log t)$ time; update operation insertEdge takes $O(\log n)$ amortized time.*

*Proof.* (*Data Structure*) We extend the data structure of [63]. We maintain the *block-cut-vertex forest* $\mathcal{B}$ of $G$. Each tree $T$ in $\mathcal{B}$ corresponds to a connected component of $G$. There are two types of nodes in $T$: *block nodes* that correspond to blocks (biconnected components) of $G$ and *vertex nodes* that correspond to vertices of $G$. Each edge of $T$ connects a vertex node to a block node. The block node associated with a block $B$ is adjacent to the vertex nodes associated with the vertices of $B$. We have that two vertices $u$ and $v$ of $G$ are in the same biconnected component if and only if there is a path between the vertex nodes of $\mathcal{B}$ associated with $u$ and $v$ and this path has length 2. Thus, operation areBiconnected in $G$ is reduced to performing operation pathLength in $\mathcal{B}$ and certifying that the returned path length equals 2.

(1) – (4) & (*Security*) They follow immediately from Theorem 3. The time complexity for edge insertions is amortized, because these are the guarantees for the data structure in [63]. □

### 4.6 Triconnectivity Queries on Graphs

Finally, we show how to authenticate the following query operation:

- areTriconnected$(u, v)$: determine whether $u$ and $v$ are in the same triconnected component of $G$,

which we call a *triconnectivity* query, in a general graph $G$ maintained through edge and vertex insertions as before. Again, we use the results of Theorem 2 and 3 to construct an authenticated data structure that answers triconnectivity queries.

**Theorem 6.** *Let $G$ be a general graph with $n$ nodes. There exists a semi-dynamic authenticated data structure that supports query operation areTriconnected on pairs of nodes in graph $G$ that evolves through update operations makeVertex and insertEdge having the following performance:*

1. *query operation areTriconnected takes $O(\log n)$ time; for this query operation, the answer authentication information has size $O(\log n)$ and the answer verification time is $O(\log n)$;*

2. *the total space used is $O(n)$;*

3. *for every update operation the update authentication information is $O(1)$;*

4. *update operation makeVertex takes $O(\log n)$ time; update operation insertEdge takes $O(\log n)$ amortized time.*

*Proof.* (*Data Structure*) We extend the data structure of [22], where a biconnected graph (or component) $G$ is associated with an *SPQR tree* $T$ that represents a recursive decomposition of $G$ by means of separation pairs of vertices. Each S-, P-, and R-node of $T$ is associated with a triconnected component $C$ of $G$ and stores a separation pair $(s, t)$, where vertices $s$ and $t$ are called the *poles* of $C$. A Q-node of $T$ is associated with an edge of $G$. Each vertex $v$ of $G$ is allocated at several nodes of $T$ and has a unique *proper allocation node* in $T$.

Our authenticated data structure augments tree $T$ with V-nodes associated with the vertices of $G$ and connects the V-node of a vertex $v$ to the proper allocation node of $v$ in $T$. Also, it uses node attributes to store the type (S, P, Q, R, or V) of a node of $T$ and its poles. In this setting, operation areTriconnected can be reduced to a small number of pathLength and type queries on the augmented SPQR tree.

$(1) - (4)$ & (*Security*) They follow immediately from Theorem 3 and the complexity bounds of the data structure in [22]. □

## 5 Authenticated Geometric Searching

In this section, we consider authenticated data structures for geometric searching problems. Such data structures have applications to the authentication of geographic information systems.

### 5.1 Fractional Cascading

Fractional cascading, originally presented in [14], is a general algorithmic technique used in a broad class of geometric data query problems. In fact, fractional cascading is an efficient strategy for solving the *iterative search* problem which is described in the sequel.

Let $U$ be an ordered universe and $\mathcal{C} = \{C_1, C_2, ..., C_k\}$ a collection of $k$ catalogs, where each catalog $C_i$ is an ordered collection of $n_i$ elements chosen from $U$. For any element $x \in U$, the *successor* of $x$ in $C_i$ is defined to be the smallest element in $C_i$ that is equal or greater than $x$. We say that we *locate* $x$ in $C_i$ when we find the successor of $x$ in $C_i$. In the iterative search problem, given an element $x \in U$, we want to locate $x$ in each catalog in $\mathcal{C}$.

Let $n = \sum_{i=1}^{k} n_i$ be the total number of stored elements. The straightforward solution is to perform $k$ separate searches: the search in catalog $C_i$ can be performed in time $O(\log n_i)$ by binary search. The total time needed is $O(k \log n)$. An alternative approach is to merge the $k$ catalogs into a master catalog $M$ and keep a correspondence dictionary between positions in $M$ and positions in each $C_i$. Using binary search on this merged catalog, we solve the problem in $O(k + \log n)$ time, but we pay the overhead of increasing the storage from $O(n)$ to $O(kn)$.

Fractional cascading succeeds in achieving an $O(k + \log n)$ time complexity for iterative search while keeping the storage linear. In comparison with the straightforward solution, we can, for instance, see that if all catalogs have the same size and $k = \sqrt{n}$ or, even $k = \log n$, we have time improvement of $O(\log n)$. The key point is how the catalog correlation that guides the search between incident catalogs is accomplished and still the storage is kept linear.

We now review the fractional cascading framework as it is usually appeared in applications. The original work in [14] covers a more general model which is mostly interesting from a theoretical point of view. We consider a static setting where catalogs are fixed and do not evolve over time. The dynamization of the technique has been studied in [41].

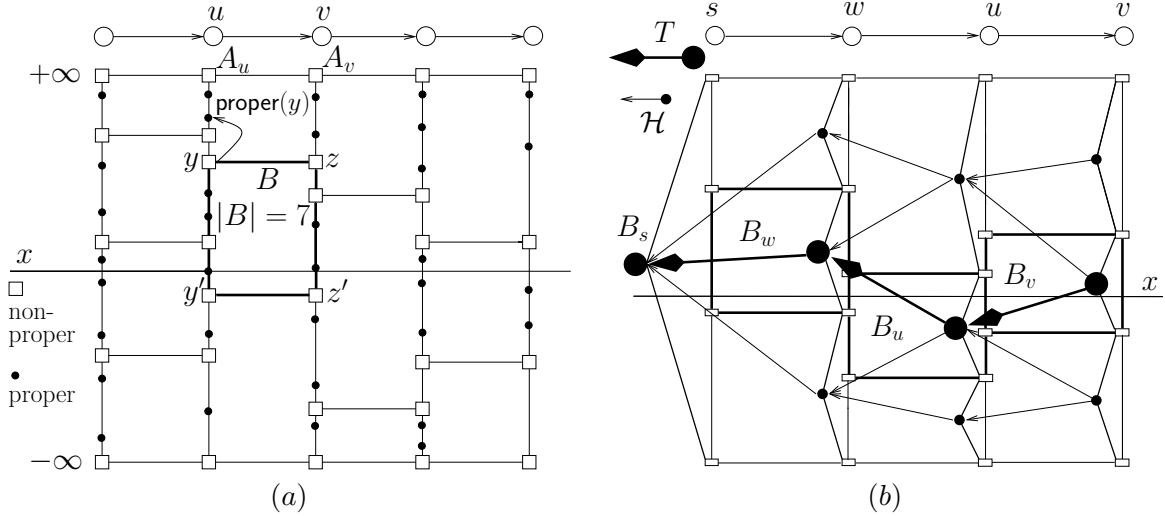

Figure 5: (a) The fractional cascading data structure over a path. Squares and dots represent non-proper and proper elements respectively. Edge $(u, v)$ has three blocks. (b) Inter-block hashing: DAG $\mathcal{H}$ defines the second-level hashing. For any query element $x$, any query graph $Q$ and any traversal of $Q$, the target blocks define a tree $T$.

Given a collection $\mathcal{C}$ of catalogs, we consider a one-to-one correspondence between the catalogs in $\mathcal{C}$ and the nodes of a graph $G$. Let $G$ be a *single source* directed acyclic graph, without multiple edges, that has bounded degree, i.e., each node of $G$ has both in-degree and out-degree bounded by a constant $d$. Each node $v$ of $G$ is associated with a catalog $C_v$. $G$ is called a *catalog* graph. Given a catalog graph $G$, we define $\mathcal{Q}(G)$ to be the family of all connected subgraphs $Q = (V, E)$ of $G$ that contain $s$ and do not contain any other node (except $s$) having zero in-degree in $Q$. The iterative search problem for the catalog graph $G$ can then be restated as: given an element $x \in U$ and a member $Q = (V, E)$ of $\mathcal{Q}(G)$, locate $x$ in $C_v$ for all $v \in V$. We refer to $Q$ as the *query* graph.

Let $k$ be the number of vertices of $G$ and let $n$ be the total number of elements in the catalogs of $G$. In [14], it is showed that using the fractional cascading technique we can build a data structure over $G$ in $O(n)$ time using $O(n)$ space and solve the iterative search problem in time $O(k + \log n)$. We will briefly describe the data structure used and the fractional cascading technique.

Each catalog $C_v$ is augmented to a catalog $A_v$ by storing some extra elements. In augmented catalog $A_v$, elements in $C_v$ are called *proper* and the other (extra) elements are called *non-proper*. Augmented catalogs that correspond to adjacent nodes of $G$ are connected via *bridges*. Let $e = (u, v)$ be an edge of $G$. A bridge connecting $A_u$ and $A_v$ is a pair $(y, z)$ associating two non-proper elements $y$ and $z$, where $y \in A_u$, $z \in A_v$ and $y = z$. Elements $y$ and $z$ have references to each other. Each non-proper element $y$ belongs to exactly one bridge. Two neighboring catalogs $A_u$ and $A_v$ are connected through at least two extreme bridges that correspond to non-proper elements $+\infty$ and $-\infty$ respectively. Each pair of neighboring bridges $(y, z)$, $(y', z')$ of edge $(u, v)$ defines a *block* $B$ which contains all elements of $A_u$ and $A_v$ lying between the two bridges. If $y' \leq y$, then bridges $(y, z)$ and $(y', z')$ are respectively called the *higher* and the *lower* bridge of $B$. The size $|B|$ of a block $B$ is the number of the elements (both proper and non-proper) that it contains. Block sizes constitute a crucial parameter for the performance of the data structure. If $n$ is the total size of the original catalogs, i.e., the total number of proper elements, then, as shown in [14], the total number of non-proper elements is $O(n)$ only if block sizes are proportional to the bounded degree $d$ of $G$. Thus, in fractional cascading blocks are chosen to have bounded size both from above and below: for each block $B$, $\alpha \leq |B| \leq \beta$, where $\alpha$ and $\beta$ are some constants proportional to $d$. Each non-proper element $z \in A_v$ is associated (by storing a reference) to its next proper element proper($z$) in $A_v$, i.e., its successor in the original catalog $C_v$. Given $z$, proper($z$) can be retrieved in constant time (e.g. $z$ points to proper($z$)). If $z$ is a proper element, then we define proper($z$) = $z$. Figure 5(a) describes the data structure built over a path $G$. We next describe how this data structure is used to locate queried elements.

## 5.2 Location Process

Suppose that we know the successor, say $l$, of $x$ in an augmented catalog $A_u$. Then we can locate $x$ at the corresponding original catalog $C_u$ in $O(1)$ time, by just retrieving proper($l$). Moreover, and if $e = (u, v) \in E$, we can locate $x$ in $A_v$ as follows: starting from $l$, we traverse $A_u$ moving to higher elements until a bridge, say $(y, z)$, is reached that connects to $A_v$, we then follow that bridge and finally traverse $A_v$ moving to smaller elements until $x$ has been located. Bridge $(y, z)$ is called the *entrance* bridge of catalog $A_u$. In this way, the search is propagated to a neighboring new node by means of a block $B$ (whose higher bridge is $(y, z)$) in time $O(|B|) = O(d) = O(1)$. Thus, having located $x$ in an original catalog, we can locate $x$ in an adjacent original catalog in constant time.

Given the query value $x$ and a query graph $Q$, we initially perform a binary search to locate $x$ at the augmented catalog $A_s$, where $s$ is the unique source node of $G$ (e.g., the root node if $G$ is a tree—the typical case in many applications) in $O(\log n)$ time. Recall that the query graph $Q$ always contains the source node $s$. Starting from node $s$, we traverse $Q$ and visit all of its nodes once. Given $Q$, such a traversal of $Q$ can be performed by considering any topological order of $Q$. A move from a node $u$ to an adjacent in $Q$ node $v$, corresponds to the procedure described above: having located $x$ in $C_u$, we locate $x$ in $C_v$ in constant time. That is, we traverse edge $(u, v)$ by moving through the entrance bridge of $A_v$. By traversing the query graph $Q$ in this way, we can solve the iterative search problem in $O(kd + \log n)$ time, where $k$ is the number of vertices of $Q$, $n$ is the total number of proper elements in the catalogs of $G$ and $d$ is the bounded degree of $G$ (a constant).

## 5.3 Hashing Scheme for Fractional Cascading

We now move to the authentication of the iterative search problem that is solved using the fractional cascading data structure. The idea is again to construct a hashing scheme over the data structure, so that a digest is computed and signed by the source and short answer authentication information is provided for any iterative search query for the catalog graph $G$. We next describe our authenticated data structure $\mathcal{D}$ for fractional cascading, which in turn is based on the path hash accumulator hashing scheme presented in Section 3.

Let $h$ is a commutative cryptographic collision-resistant hash function. We assume that a set of rules have been defined, so that $h$ can operate on elements of catalogs, nodes of graph $G$ and previously computed hash values. The hashing scheme can be viewed as a two-level hashing structure, built using the path hash accumulator scheme: *intra-block* hashing is performed within each block defined in the data structure and *inter-block* hashing of performed through all blocks of the data structure. In the sequel, we describe each hashing structure.

**Intra-block hashing:** Consider any edge $(u,v)$ of $G$, i.e., $u$ is one of the parents (i.e., predecessor nodes) of $v$, and the corresponding augmented catalogs $A_u$ and $A_v$. Also, consider any two neighboring bridges $(y',z')$ and $(y,z)$ connecting $A_u$ and $A_v$ that define block $B$. Assume that $z,z' \in A_v$. We define $P$ to be the sequence of elements of $B$ that exist in $A_v$ plus the non-proper elements of the corresponding bridges that lie in $A_v$. That is, $P = \{p_1, p_2, ..., p_t\}$ is an increasing sequence, where, if $z' \leq z$, $p_1 = z'$ and $p_t = z$. We refer to $P$ as the *hash side* of $B$. Using the path hash accumulator scheme, we compute the digest $D(P)$ of sequence $P$. For each element $p_i$, we set $\mathcal{N}(p_i) = \{p_i, \mathsf{proper}(p_i), v\}$ and in that way the path hash accumulator can support authenticated membership queries and authenticated path property queries. Here, one property of $P$ is the corresponding node $v$.

We iterate the process for all blocks defined in the data structure: for each block $B$ having a hash side $P$ in $A_v$, $H_B$ is the hash path accumulation $D(P)$ of sequence $P$. We also define $B_s$ to be a fictitious block, the augmented catalog $A_s$. The hash side of $B_s$ is the whole block itself, so $H_{B_s}$ is well defined. All the path hash accumulators used define the first-level hashing structure.

**Inter-block hashing:** The second-level hashing structure is defined through a directed acyclic graph $\mathcal{H}$ defined over blocks. In particular, nodes of $\mathcal{H}$ are blocks of the data structure. Suppose that $w$ is a parent (i.e., predecessor node) of $u$ and $u$ is a parent of $v$ in $G$. If $B$ is a block of edge $(u,v)$, then we add to the set of edges of $\mathcal{H}$ all the directed edges $(B, B')$, where $B'$ is a block of edge $(w,u)$ that shares elements from $A_u$ with $B$. Additionally, if $v$ is a child (i.e., successor node) of the root (i.e., unique source node) $s$ in $G$, then for all blocks $B$ that correspond to edge $(s,v)$, we add to the set of edges of $\mathcal{H}$ the directed edge $(B, B_s)$. This completes the definition of graph $\mathcal{H}$. Note that $B_s$ is the unique sink node of $\mathcal{H}$. Figure 5(b) shows the graph $\mathcal{H}$ that corresponds to a catalog graph $G$ that is simply a path.

Each block (node) $B$ of $\mathcal{H}$ is associated with a label $L(B)$. If $B$ is a source node (leaf) in $\mathcal{H}$, then $L(B) = H_B$. If $B$ is the successor node (parent) of blocks $B_1, \ldots, B_t$ in $\mathcal{H}$, listed in some fixed order, then $L(B)$ equals the path hash accumulation over sequence $B_1, \ldots, B_t$ using $\mathcal{N}(B_i) = \{L(B_i), H_B\}$. We emphazise that in this case (i.e., for the inter-block hashing) the path hash accumulator is *merely* used for authenticating membership in a set (like a Merkle tree; thus, no path property is used). This hashing scheme over $\mathcal{H}$ corresponds to the second-level hashing structure.

Finally, we set $D(\mathcal{D}) = L(B_s)$ to be the digest of the entire data structure $\mathcal{D}$ (which is signed by the data source).

## 5.4 Answer Authentication Information

Given a query $x$ and a query graph $Q$, we describe now what is the authentication information given to the user. If $v$ is a node of $Q$, let $s_v$ be the successor of $x$ in $C_v$. In the location process, to locate $x$ in $A_v$, we find two consecutive elements $y$ and $z$ of $A_v$ such that $y \leq x < z$, where each of $y$ and $z$ may be either proper or non-proper. They are both elements of a block $B$ such that the entrance bridge of $A_v$ is the higher bridge of $B$. Observe that $z$ is the successor of $x$ in $A_v$ and that $s_v = \mathsf{proper}(y)$ when $y = x$, or $s_v = \mathsf{proper}(z)$ when $y < x$. We call $z$ and $B$, the *target element* and the *target block* of $A_v$, respectively.

Two useful observations are that: (1) in the location process, the traversal of the query graph $Q$ is chosen so that each node of $Q$ is visited once and (2) any two target blocks visited by the location process that correspond to incident edges in $Q$ share elements of the common augmented catalog, and, thus, are adjacent in graph $\mathcal{H}$. It follows that all the target blocks define a subgraph $T$ of $\mathcal{H}$. This subgraph $T$ consists of the all target blocks and the edges of $\mathcal{H}$ that connect neighboring target blocks (Figure 5(b)).

**Lemma 7.** *For any query graph $Q$, graph $T$ is a tree.*

*Proof.* Consider the topological order used to define the traversal of the query graph $Q$. This topological order defines a directed subtree $T_Q$ of $Q$. There is an one-to-one correspondence between edges of $T_Q$ and target blocks, i.e., between edges of $T_Q$ and nodes of $T$. $\square$

For any node $v$, let $z_v$ be the target element of $A_v$ and $B_v$ the target block of $A_v$. Then, the answer authentication information for our hashing scheme consists of:

1. *Intra-block:* for each node $v$ of $Q$, the target element $z_v$ of $A_v$ and a verification sequence $p_v$ from $z_v$ up to the path hash accumulation of the hash side of $B_v$, and

2. *Inter-block:* for every node (or target block) $B_v$ of $T$ that is not a leaf, the verification sequences from every child of $B_v$ in $T$ up to the path hash accumulation $L(B_v)$.

**Lemma 8.** *If $n$ is the total number of proper elements in the catalogs of $\mathcal{C}$ and $d$ is the bounded degree of $G$, then for any query graph $Q$ of $k$ nodes, the size of the answer authentication information is $O(\log n + k \log d) = O(\log n + k)$.*

*Proof.* The hash side of $B_s$ has size $|A_s| = O(n)$ and the hash side of any other target block has size $O(d)$. Thus, the intra-block answer authentication information consists of $k$ verification sequences: $k - 1$ of size $O(\log d)$ and one of size $O(\log n)$. Therefore, this portion of the answer authentication information has $O(\log n + k \log d)$ size. For the inter-block answer authentication information, recall that $G$ and, thus, both $Q$ and $T_Q$, have out-degree bounded by $d$ and that every target block can share elements with at most $O(d)$ other target blocks. Thus, $\mathcal{H}$ and $T$ have in-degree bounded by $O(d)$. Now, all, but $L(B_s)$, the second-level path hash accumulations are built over sequences of length $O(d)$. Path hash accumulation $L(B_s)$ is built over at most $dn$ blocks that share elements with $A_s$. Observe that there is a one-to-one correspondence between inter-block verification sequences and edges in $T$. It follows that the inter-block answer authentication information consists of $k - 2$ verification sequences of size $O(\log d)$ and one of size $O(\log n)$, thus, this second portion of the answer authentication information has also $O(\log n + k \log d)$ size. In total, since $d$ is a constant, the answer authentication information is of size $O(\log n + k)$. $\square$

### 5.5 Answer Verification

For a given query element $x$ and query graph $Q = (V, E)$, we assume that the answer given to the user is a set $A = \{(a_v, v) : v \in V\}$, where $a_v$ is claimed to be the successor of $x$ in $C_v$. The answer authentication information consists of two verification sequences for each node (target block) of tree $T$: one intra-block and one inter-block. These sequences form a hash tree in our two-level hashing scheme. The verification process is defined by this hash tree. Intuitively, an intra-block verification sequence of a target block $B_v$ provides a *local proof* that $a_v$ is the successor of $x$ in $C_v$, and then, all these local proofs are accumulated through inter-block verification sequences into the (signed) digest.

Given elements $x$, $y$, $y'$, $z$, $z'$ and a node $v$ of $Q$, consider the predicates (or, alternatively, relations): (1) $y \leq x < z$, (2) $y$ and $z$ are consecutive elements in $A_v$, (3) $y = x$ and $y' = \mathsf{proper}(y)$ in $A_v$ and (4) $y < x$ and $z' = \mathsf{proper}(z)$ in $A_v$. If (1), (2) and (3) hold simultaneously, then they constitute a proof that the successor of $x$ in $C_v$ is $y'$, whereas if (1), (2) and (4) hold simultaneously, they constitute a proof that the successor of $x$ in $C_v$ is $z'$. Such a proof must be provided for every $v$ of $Q$.

Given $A$, $x$ and the answer authentication information, the user first checks if there is any inconsistency between values $a_v$ and $z_v$ for every $v$ of $Q$ with respect to the two possible proofs above. Observe that, by the answer authentication information, the user knows for each node $v$ of $Q$ the target element $z_v$ and the corresponding element $y_v$, such that $y_v < z_v$ and $y_v$ and $z_v$ are consecutive elements in $A_v$. If there is at least one inconsistency, the user rejects the answer. Otherwise, all that is needed is to verify the signed digest $D(\mathcal{D})$ of the data structure. Observe, that the user possesses all the data needed for the recomputation of the signed digest. If the computed digest matches the verified signed digest, then based on the collision-resistance property of the hash function used in the scheme, the user has a proof that the answer is correct and accordingly accepts the answer as authentic. Otherwise (if either the signed digest is not verified or the recomputed digest does not match the signed one), the user rejects the answer as invalid.

**Lemma 9.** *If $n$ is the total number of proper elements in the catalogs of $\mathcal{C}$, then for any query graph $Q$ of $k$ nodes, the answer verification time is $O(\log n + k \log d) = O(\log n + k)$, where $d$ is the bounded degree of $G$.*

*Proof.* It follows directly from Lemma 8. Recall that the verification time of a path hash accumulator is proportional to the size of the verification sequence. □

If the digest is verified, then based on the collision-resistance property of the hash function $h$, and, in particular, the security of the path hash accumulator, the user has a proof that the answer is correct: for each $v$ of $Q$, the user can verify all the three conditions previously discussed. A faulty answer can lead to a forged proof only if some collisions of $h$ have been found: the responder needs to break the security of the path hash accumulator in authenticating membership queries, which is further reduced to finding collisions of the cryptographic hash function $h$ in use.

**Lemma 10.** *For any catalog graph $G$ of $k$ nodes and of total size $n$, both intra-block and inter-block hashing schemes can be computed in $O(n)$ time using $O(n)$ storage.*

*Proof.* $G$ has in-degree that is bounded by $d = O(1)$ and every target block can share elements with at most $\beta = O(1)$ other blocks. Moreover, the path hash accumulation of a sequence of length $m$ can be computed in $O(m)$ time and space. □

We have thus proved the following theorem.

**Theorem 11.** *Let $G$ be the catalog graph for a collection $\mathcal{C}$ of $t$ catalogs and $n$ be the total number of elements stored in $\mathcal{C}$, where $t \leq n$. If $G$ is of bounded degree, then the authenticated fractional cascading data structure $\mathcal{D}$ for $G$ solves the authenticated iterative search problem for $G$, achieving the following performance:*

1. *$\mathcal{D}$ can be constructed in $O(n)$ time and uses $O(n)$ storage;*

2. *given a query element $x$ and a query graph $Q$ with $k \leq t$ vertices, $x$ can be located in every catalog of $Q$ in $O(\log n + k)$ time;*

3. *the answer authentication information has size $O(\log n + k)$ and the answer verification time is $O(\log n + k)$.*

## 5.6 Applications of Authenticated Iterative Search

Our authenticated fractional cascading scheme can be used to design authenticated data structures for various fundamental two-dimensional geometric search problems, where iterative search is implicitly performed (see [15]). In all of these problems, the underlying catalog graph has degree bounded by a small constant, and in most cases the graph itself is a tree. We next describe how the authentication of the iterative search problem can be extended to provide authentication of this broad class of queries that involves searching in multi-catalogs that are organized in a tree.

**Authentication scheme.** The idea, here, is to extend the hashing scheme of the fractional cascading data structure over the graph structure in which catalogs are organized. In essence, what we need to additionally authenticate is that the correct subdigraph of the catalog graph is accessed by the responder and used to generate the answer. That is, so far (for the iterative search only) we assumed that this catalog subgraph is part of the query (query graph $Q$). In the applications of the fractional cascading data structure, however, this subgraph is not known in advance, but it is rather generated on-the-fly as the answer is being produced. In other words, the query graph is in essence part of the answer itself. Accordingly, in order to verify the answer, the user needs first to authenticate that the correct (authentic) subgraph is generated by the responder and that the final answer corresponds to this correct subgraph of catalogs. Given that in our authentication schemes verification is performed in a bottom-up fashion, the user essentially will first authenticate the iterative search (as described in the previous subsections), and then he will authenticate that the subgraph that corresponds to the iterative search is authentic (all nodes that should have been included in the graph have been included, but no extra nodes are included). This second verification step (subgraph authentication) can be easily performed as follows.

The catalog graph is accessed by some search algorithm. The key property that we wish our authentication scheme to satisfy is to authenticate this search procedure. This is done by hashing over the graph structure (bottom-up hashing over the graph) as follows. Let $v$ be a node of the catalog graph $G$—recall, $G$ is a directed acyclic graph—that has $u_1, \ldots, u_\ell$ successor nodes, where $\ell$ is a constant (because the catalog graph $G$ is of bounded degree). Let $d_v$ be the data that is used by the search algorithm to advance the search from node $v$ to one or more successor nodes. Typically, especially for catalogs organized as trees, the search is performed over an ordered data set, and, in this case, $d$ is a sequence of $\ell$ (or $\ell - 1$) keys (members of the set) that are used along with the search items (defined by the query) to decide in which node(s) to advance the search. Node $v$ is then associated with a hash value $h_v$, defined as

$$h_v = h(h_{u_1}, \ldots, h_{u_\ell}, h(d_v)).$$

Thus, we have a recursive definition of the digest of the hashing scheme at the source of $G$. Depending on exact format of data $d_v$, additional improvements (up to some constant) may be considered in producing the hash value (digest) $h(d_v)$. Given, a traversed subgraph of $G$, now, the answer authentication information additionally contains the hash values and search information (i.e., keys) that are needed to recompute the final data digest. Given this authentication scheme, the verification of the answer first involves the verification of the iterative search problem, as described earlier in the section, and then the verification of the search subgraph of catalogs (depending on the exact query problem, the order may be swapped). The connection of the two individual hashing schemes can be easily performed at the root of the entire hashing structure: simply by hashing together the two individual digests to generate the digest of the entire hashing scheme.

Given this extra layer of hashing and this extension of our authentication scheme, we obtain authenticated versions of any data structure that uses iterative search over set of catalogs organized as nodes of a DAG. The following results are obtained by using the results in [15] and the results of Theorem 11 for our authentication schemes of this section, where $n$ denotes the problem size.

**Corollary 12.** *There is an authenticated data structure for answering line-intersection queries on a polygon with $n$ vertices that can be constructed in $O(n \log n)$ time and uses $O(n \log n)$ storage. Denoting with $k$ the output size, queries are answered in $O(\log n + k)$ time; the answer authentication information has size $O((k+1) \log \frac{n}{k+1})$; and the answer verification time is $O((k+1) \log \frac{n}{k+1})$.*

**Corollary 13.** *There are authenticated data structures for answering ray shooting and point location queries on a planar subdivision with $n$ vertices that can be constructed in $O(\log n)$ time and use $O(n \log n)$ storage. Queries are answered in $O(\log n)$ time; the answer authentication information has size $O(\log n)$; and the answer verification time is $O(\log n)$.*

**Corollary 14.** *There are authenticated data structures for answering orthogonal range search, orthogonal point enclosure and orthogonal intersection queries that can be constructed in $O(n \log n)$ time and use $O(n \log n)$ storage, where $n$ is the problem size. Denoting with $k$ the output size, queries are answered in $O(\log n + k)$ time; the answer authentication information has size $O(\log n + k)$; and the answer verification time is $O(\log n + k)$.*

# 6 Conclusion

In this paper, we have examined the problem of designing efficient authenticated data structures for broad classes of queries. We have developed the path hash accumulator, a new authentication scheme for general decomposable queries over sequences of data elements and, in particular, queries about properties of subsequences that involve any associative operation applied over their elements. Using this authentication scheme, we then design new authenticated data structures for graph queries (e.g., path and connectivity queries) or search problems over two-dimensional geometric objects (e.g., point location and range search). Authentication of graph queries is performed by authenticating certain path properties in some tree graphs that are specially designed for the graph in question. Authentication of geometric search problems is performed by authenticating the general static version of the fractional cascading framework that solves the iterative search problem. Our authentication techniques are efficient and introduce asymptotically no extra overhead to the underlying search structure.

An interesting open problem is the design of dynamic versions of our authenticated data structures based on fractional cascading.

## Acknowledgments

Research supported in part by NSF grants CCF–0311510, IIS–0324846, IIS–0713046, IIS–0713403, OCI–0724806, and CCF–0830149, and by a research gift from Sun Microsystems.

We would like to thank Robert Cohen for contributing to the results of Section 4 and for coauthoring a preliminary version of this paper [33]. We also thank the anonymous reviewers of this work for their useful comments.

# References

[1] W. Aiello, S. Lodha, and R. Ostrovsky. Fast digital identity revocation. In *Advances in Cryptology—CRYPTO*, volume 1462 of *LNCS*, pages 137–152. Springer, 1998.

[2] A. Anagnostopoulos, M. T. Goodrich, and R. Tamassia. Persistent authenticated dictionaries and their applications. In *Proc. Inf. Security Conf.*, volume 2200 of *LNCS*, pages 379–393. Springer, 2001.

[3] M. J. Atallah, Y. Cho, and A. Kundu. Efficient data authentication in an environment of untrusted third-party distributors. In *Proc. Int. Conf. on Data Eng.*, pages 696–704. IEEE, 2008.

[4] N. Baric and B. Pfitzmann. Collision-free accumulators and fail-stop signature schemes without trees. In *Advances in Cryptology—EUROCRYPT*, volume 1233 of *LNCS*, pages 480–494. Springer, 1997.

[5] J. Benaloh and M. de Mare. One-way accumulators: A decentralized alternative to digital signatures. In *Advances in Cryptology—EUROCRYPT*, volume 765 of *LNCS*, pages 274–285. Springer, 1993.

[6] S. W. Bent, D. D. Sleator, and R. E. Tarjan. Biased search trees. *SIAM J. Comput.*, 14:545–568, 1985.

[7] E. Bertino, B. Carminati, E. Ferrari, B. Thuraisingham, and A. Gupta. Selective and authentic third-party distribution of XML documents. *IEEE Trans. Knowledge and Data Eng.*, 16(10):1263–1278, 2004.

[8] M. Blum and S. Kannan. Designing programs that check their work. *J. ACM*, 42(1):269–291, 1995.

[9] J. D. Bright and G. Sullivan. Checking mergeable priority queues. In *Digest Symp. on Fault-Tolerant Comput.*, pages 144–153. IEEE, 1994.

[10] J. D. Bright and G. Sullivan. On-line error monitoring for several data structures. In *Digest Symp. on Fault-Tolerant Comput.*, pages 392–401. IEEE, 1995.

[11] J. D. Bright, G. Sullivan, and G. M. Masson. Checking the integrity of trees. In *Digest Symp. on Fault-Tolerant Comput.*, pages 402–411. IEEE, 1995.

[12] A. Buldas, P. Laud, and H. Lipmaa. Eliminating counterevidence with applications to accountable certificate management. *J. Computer Security*, 10(3):273–296, 2002.

[13] J. Camenisch and A. Lysyanskaya. Dynamic accumulators and application to efficient revocation of anonymous credentials. In *Advances in Cryptology—CRYPTO*, volume 2442 of *LNCS*. Springer, 2002.

[14] B. Chazelle and L. J. Guibas. Fractional cascading: I. A data structuring technique. *Algorithmica*, 1(3):133–162, 1986.

[15] B. Chazelle and L. J. Guibas. Fractional cascading: II. Applications. *Algorithmica*, 1:163–191, 1986.

[16] R. F. Cohen and R. Tamassia. Combine and conquer. *Algorithmica*, 18:342–362, 1997.

[17] P. Devanbu, M. Gertz, A. Kwong, C. Martel, G. Nuckolls, and S. Stubblebine. Flexible authentication of XML documents. *J. Computer Security*, 6:841–864, 2004.

[18] P. Devanbu, M. Gertz, C. Martel, and S. G. Stubblebine. Authentic data publication over the

Internet. *J. Computer Security*, 11(3):291–314, 2003.

[19] O. Devillers, G. Liotta, F. P. Preparata, and R. Tamassia. Checking the convexity of polytopes and the planarity of subdivisions. *Comput. Geom. Theory Appl.*, 11:187–208, 1998.

[20] G. Di Battista and G. Liotta. Upward planarity checking: "Faces are more than polygons". In S. H. Whitesides, editor, *Proc. Graph Drawing*, volume 1547 of *LNCS*, pages 72–86. Springer, 1998.

[21] G. Di Battista and B. Palazzi. Authenticated relational tables and authenticated skip lists. In *Proc. IFIP Conf. on Database Security*, volume 4602 of *LNCS*, pages 31–46. Springer, 2007.

[22] G. Di Battista and R. Tamassia. On-line maintenance of triconnected components with SPQR-trees. *Algorithmica*, 15:302–318, 1996.

[23] D. Eppstein, G. F. Italiano, R. Tamassia, R. E. Tarjan, J. Westbrook, and M. Yung. Maintenance of a minimum spanning forest in a dynamic plane graph. *J. Algorithms*, 13(1):33–54, 1992.

[24] U. Finkler and K. Mehlhorn. Checking priority queues. In *Proc. Symp. on Discrete Algorithms*, pages 901–902. SIAM, 1999.

[25] I. Gassko, P. S. Gemmell, and P. MacKenzie. Efficient and fresh certification. In *Proc. Int. Conf. on Pract. and Theory in Public Key Cryptography*, volume 1751 of *LNCS*, pages 342–353. Springer, 2000.

[26] S. Goldwasser, S. Micali, and R. Rivest. A digital signature scheme secure against adaptive chosen-message attacks. *SIAM J. Comput.*, 17(2):281–308, 1988.

[27] M. T. Goodrich, C. Papamanthou, and R. Tamassia. On the cost of persistence and authentication in skip lists. In *Proc. Int. Workshop on Experimental Algorithms*, volume 4525 of *LNCS*, pages 94–107. Springer, 2007.

[28] M. T. Goodrich, C. Papamanthou, R. Tamassia, and N. Triandopoulos. Athos: Efficient authentication of outsourced file systems. In *Proc. Inf. Security Conf.*, volume 5222 of *LNCS*, pages 80–96. Springer, 2008.

[29] M. T. Goodrich and R. Tamassia. Efficient authenticated dictionaries with skip lists and commutative hashing. Technical report, Johns Hopkins Information Security Institute, 2000. Available from http://www.cs.brown.edu/cgc/stms/papers/hashskip.pdf.

[30] M. T. Goodrich, R. Tamassia, and J. Hasic. An efficient dynamic and distributed cryptographic accumulator. In *Proc. Inf. Security Conf.*, volume 2433 of *LNCS*, pages 372–388. Springer, 2002.

[31] M. T. Goodrich, R. Tamassia, and A. Schwerin. Implementation of an authenticated dictionary with skip lists and commutative hashing. In *Proc. DARPA Inf. Survivability Conf. and Exposition*, volume 2, pages 68–82, 2001.

[32] M. T. Goodrich, R. Tamassia, and N. Triandopoulos. Super-efficient verification of dynamic outsourced databases. In *Proc. RSA Conf., Cryptographers' Track*, volume 4964 of *LNCS*, pages 407–424. Springer, 2008.

[33] M. T. Goodrich, R. Tamassia, N. Triandopoulos, and R. Cohen. Authenticated data structures for graph and geometric searching. In *Proc. RSA Conf., Cryptographers' Track*, volume 2612 of *LNCS*, pages 295–313. Springer, 2003.

[34] A. Heitzmann, B. Palazzi, C. Papamanthou, and R. Tamassia. Efficient integrity checking

of untrusted network storage. In *Proc. Int. Workshop on Storage Security and Survivability*, pages 43–54. ACM, 2008.

[35] V. King. A simpler minimum spanning tree verification algorithm. In *Proc. Int. Workshop on Algorithms and Data Structures*, volume 955 of *LNCS*, pages 440–448. Springer, 1995.

[36] P. C. Kocher. On certificate revocation and validation. In *Proc. Int. Conf. on Financial Cryptography*, volume 1465 of *LNCS*, pages 172–177. Springer, 1998.

[37] F. Li, M. Hadjieleftheriou, G. Kollios, and L. Reyzin. Dynamic authenticated index structures for outsourced databases. In *Proc. Int. Conf. on Management of Data*, pages 121–132. ACM, 2006.

[38] J. Li, N. Li, and R. Xue. Universal accumulators with efficient nonmembership proofs. In *Proc. Int. Conf. on Applied Cryptography and Network Security*, volume 4521 of *LNCS*, pages 253–269, 2007.

[39] P. Maniatis and M. Baker. Secure history preservation through timeline entanglement. In *Proc. USENIX Security Symp.*, pages 297–312. USENIX, 2002.

[40] C. Martel, G. Nuckolls, P. Devanbu, M. Gertz, A. Kwong, and S. G. Stubblebine. A general model for authenticated data structures. *Algorithmica*, 39(1):21–41, 2004.

[41] K. Mehlhorn and S. Näher. Dynamic fractional cascading. *Algorithmica*, 5(1–4):215–241, 1990.

[42] K. Mehlhorn and S. Näher. *LEDA: A Platform for Combinatorial and Geometric Computing*. Cambridge University, Cambridge, UK, 2000.

[43] K. Mehlhorn, S. Näher, M. Seel, R. Seidel, T. Schilz, S. Schirra, and C. Uhrig. Checking geometric programs or verification of geometric structures. *Comput. Geom. Theory Appl.*, 12(1–2):85–103, 1999.

[44] R. C. Merkle. A certified digital signature. In *Advances in Cryptology—CRYPTO*, volume 435 of *LNCS*, pages 218–238. Springer, 1989.

[45] S. Micali, M. Rabin, and J. Kilian. Zero-Knowledge sets. In *Proc. Symp. on Foundations of Computer Science*, pages 80–91. IEEE, 2003.

[46] M. Naor and K. Nissim. Certificate revocation and certificate update. In *Proc. USENIX Security Symp.*, pages 217–228. USENIX, 1998.

[47] M. Narasimha and G. Tsudik. Authentication of outsourced databases using signature aggregation and chaining. In *Proc. Int. Conf. on Database Systems for Advanced Applications*, volume 3882 of *LNCS*, pages 420–436. Springer, 2006.

[48] L. Nguyen. Accumulators from bilinear pairings and applications. In *Proc. RSA Conf., Cryptographers' Track*, volume 3376 of *LNCS*, pages 275–292. Springer, 2005.

[49] G. Nuckolls. Verified query results from hybrid authentication trees. In *Proc. IFIP Conf. on Database Security*, volume 3654 of *LNCS*, pages 84–98. Springer, 2005.

[50] R. Ostrovsky, C. Rackoff, and A. Smith. Efficient consistency proofs for generalized queries on a committed database. In *Proc. Int. Colloquium on Automata, Languages and Programming*, volume 3142 of *LNCS*, pages 1041–1053. Springer, 2004.

[51] H. Pang, A. Jain, K. Ramamritham, and K.-L. Tan. Verifying completeness of relational query results in data publishing. In *Proc. Int. Conf. on Management of Data*, pages 407–418. ACM, 2005.

[52] C. Papamanthou and R. Tamassia. Time and space efficient algorithms for two-party authenticated data structures. In *Proc. Int. Conf. on Inf. and Commun. Security*, volume 4861 of *LNCS*, pages 1–15. Springer, 2007.

[53] C. Papamanthou, R. Tamassia, and N. Triandopoulos. Authenticated hash tables. In *Proc. Conf. on Comput. and Commun. Security*, pages 437–448. ACM, 2008.

[54] W. Pugh. Skip lists: a probabilistic alternative to balanced trees. *Commun. ACM*, 33(6):668–676, 1990.

[55] D. D. Sleator and R. E. Tarjan. A data structure for dynamic trees. *J. Comput. Syst. Sci.*, 26(3):362–381, 1983.

[56] G. F. Sullivan and G. M. Masson. Certification trails for data structures. In *Digest Symp. on Fault-Tolerant Comput.*, pages 240–247. IEEE, 1991.

[57] G. F. Sullivan, D. S. Wilson, and G. M. Masson. Certification of computational results. *IEEE Trans. Comput.*, 44(7):833–847, 1995.

[58] R. Tamassia. Authenticated data structures. In *Proc. European Symp. on Algorithms*, volume 2832 of *LNCS*, pages 2–5. Springer, 2003.

[59] R. Tamassia and N. Triandopoulos. Computational bounds on hierarchical data processing with applications to information security. In *Proc. Int. Colloquium on Automata, Languages and Programming*, volume 3580 of *LNCS*, pages 153–165. Springer, 2005.

[60] R. Tamassia and N. Triandopoulos. Certification and authentication of data structures, 2007. Manuscript. Available at `http://www.cs.brown.edu/cgc/stms/papers/cads.pdf`.

[61] R. Tamassia and N. Triandopoulos. Efficient content authentication in peer-to-peer networks. In *Proc. Int. Conf. on Applied Cryptography and Network Security*, volume 4521 of *LNCS*, pages 354–372. Springer, 2007.

[62] R. E. Tarjan. *Data Structures and Network Algorithms*, volume 44 of *CBMS-NSF Regional Conference Series in Applied Mathematics*. SIAM, 1983.

[63] J. Westbrook and R. E. Tarjan. Maintaining bridge-connected and biconnected components on-line. *Algorithmica*, 7:433–464, 1992.